\documentclass[aps,a4paper,amsfonts,amsmath,amssymb,nofootinbib]{revtex4}
\usepackage{makeidx}
\usepackage{graphicx}
\usepackage{epsfig}
\usepackage{xcolor}

\usepackage[normalem]{ulem}

\newcommand \beq {\begin{equation}}
\newcommand \eeq {\end{equation}}
\newcommand \bd {\begin{displaymath}}
\newcommand \ed {\end{displaymath}}
\newcommand \beqn {\begin{eqnarray}}
\newcommand \eeqn {\end{eqnarray}}

\newcommand{\Erf}{\mbox{Erf}}

\newcommand{\Exp}{\mbox{Exp}}

\newcommand{\s}{\theta}
\newcommand{\x}{\zeta}
\newcommand{\sgn}{\mbox{sgn}}

\begin{document}

\title{The effect of stochastic resettings on the counting of level crossings for inertial random processes}

\author{Miquel Montero}
\email{miquel.montero@ub.edu}
\author {Matteo Palassini}
\email{palassini@ub.edu}
\author{Jaume Masoliver}
\email{jaume.masoliver@ub.edu}
\affiliation{Department of Condensed Matter Physics and Institute of Complex Systems (UBICS), \\
University of Barcelona, Catalonia, Spain}

\date{\today}

\begin{abstract}

We study the counting of level crossings for inertial random processes exposed to stochastic resetting events.  We develop the general approach of stochastic resetting for inertial processes with sudden changes in the state characterized by position and velocity. We obtain the level-crossing intensity in terms of that of underlying reset-free process, for resetting events with Poissonian statistics. We apply this result to the random acceleration process and the inertial Brownian motion. 
In both cases, we show that there is an optimal resetting rate that maximizes the crossing intensity,
and we obtain the asymptotic behavior of the crossing intensity for large and small resetting rates.
Finally, we discuss the stationary distribution and the mean first-arrival time in the
presence of resetting. 
\end{abstract}


\maketitle

\section{Introduction}
\label{intro}

The study of level crossings constitutes a very relevant aspect of stochastic processes because of its theoretical importance \cite{redner_book,maso_llibre,majumdar_pr_2020} but also for its wide practical interest \cite{blake,rychlik,lindgren_book}. The subject is diverse and embraces topics such as first-passage times, survival, escape and the theory of extremes --among many others. From a general point of view the problem consists in obtaining the probability distribution of the interval between successive crossing events, a question with no known exact solution \cite{blake,munakata} and few approximate results, mostly for Gaussian processes \cite{blake,lindgren_2019} and the random telegraph signal \cite{rickard}. 

A more accessible problem is the counting of level crossings  and the first developments along this direction were obtained in the 1940's,  especially by S. O. Rice  \cite{rice} in the context of statistical communication theory, and partly by M. Kac \cite{kac} for a purely mathematical problem (counting zeros of random polynomials), although both  approaches were limited to Gaussian and stationary random processes. The counting problem was later set on more rigorous mathematical basis by a series of authors, especially H. Cramers and collaborators \cite{cramer,cramer_leadbetter,leadbetter_spaniolo,lindgren_book,lindgren_2019}. More recently, the problem 
has been generalized to scalar-valued  \cite{longuet} and vector-valued \cite{azais,berzin} Gaussian random fields.

In engineering, applications of the level-crossing problem
range from communications, reliability and signal processing to oceanography, just to name a few \cite{rychlik,lindgren_book,lindgren_2019}. Applications in physics include persistence and first-passage properties \cite{bray_etal,maso_pala}, and the stochastic evolution of spin systems \cite{paul1, paul2}. 
A generalization of Rice's theory has also been used to count the number of critical points in stochastic processes and random fields, notably in the statistical physics of disordered systems \cite{fyodorov_04, fyodorov_05, bray_07}. 

In the counting of level crossings the main attainable result is the crossing intensity, or crossing frequency, that can be defined as the average number of times a random process intersects some given level per unit time. As shown by Rice \cite{rice} and Kac \cite{kac} the crossing intensity depends on the joint probability density function (PDF) of the process and its derivative (sometimes called ``velocity''). This joint density  is rather difficult to obtain for non Gaussian processes, and even for Gaussian processes it may not exist, as is the case, for instance, of first-order processes driven by white noise, in which the velocity has infinite variance (see the next section and Ref. \cite{maso_pala} for more details). 

In a recent work \cite{maso_pala}, two of us have studied the crossing intensity for linear second-order (i.e., inertial) processes driven by Gaussian white noise and have 
 generalized Rice's formula for such processes. Herein we address the effects of stochastic resetting events on linear inertial processes. 
In its most common form, stochastic resetting consists in the combination of a given random process with a resetting mechanism which at random instants of time (usually Poissonian) instantaneously brings the process to a given fixed position. The random dynamics of the process and the resetting mechanism are taken to be independent of each other.

Following a few antecedents in physics (e.g., \cite{eliazar_2007,manrubia}) and in the mathematics literature  (see \cite{mont_epj_2017} for more information), during the last decade there has been an explosion of works dedicated to stochastic resetting in its multiple forms and generalizations, which has given rise to a large literature starting with the work of Evans, Majumdar and collaborators \cite{majumdar_11prl,majumdar_11jpa,majumdar_13jpa,majumdar_13pre,majumdar_14,pal_2016,evans_20} as well as many others (see, for instance, the sample \cite{MV_2013,MV_2016,novak_15,belan_18,reuveni_16,biro,pal_15,Christou_15,maso_2019,maso_mont_2019,wang_22,Meylahn_2015,Harris_2017,rotbart_15,eule_16,
nagar_16,pal_reuveni_17,sokolov_18, VM_18,maso_mendez_19} out of a huge list of diverse works on the subject). 

The importance of resetting is based on two fundamental facts. Firstly, resetting stabilizes the underlying random process, in the sense that a nonstationary process becomes stationary after a resetting mechanism has been implemented. Secondly, and perhaps more importantly, resetting may reduce the mean first-arrival time, thus speeding up stochastic search algorithms. This motivates its interest 
in a variety of contexts such as protein identification in DNA sequences \cite{eliazar_2007,eliazar_2008,badr_15,reuveni_14,roldan_16}, animal foraging  \cite{mendez_13,reynolds}, internet search algorithms, data mining \cite{luby_93,montanari_02,tong_06}, and economy \cite{mpm_22}, among others.

The main objective of the present paper is to elucidate how a stochastic resetting mechanism can alter the dynamics of level crossings, especially their counting, of an underlying (reset-free) random process. In order to apply Rice's theory of crossing counting we must assume that the underlying process is of bounded variation, so that the joint PDF of position and velocity exists. This rules out random processes governed by first-order stochastic differential equations driven by white noise, as mentioned above. We will, therefore, base our study on underlying processes driven by second-order differential equations, that is inertial processes, for which velocity is bounded and the joint PDF exists. As to the resetting mechanism, we will assume that it is independent of the reset-free process and that the random instants of time at which resetting occurs are Poissonian. One important result of this paper is that, in the cases we considered and likely for all inertial processes, the crossing intensity tends to a stationary value for large times, and that this value displays a maximum as a function of the resetting rate. In other words, the average time between two consecutive crossings of a given level (the so-called return time) is minimized by an optimal choice of the resetting rate. This is reminiscent of the minimum of the mean first-absorption time in diffusive processes and other stochastic processes with resetting \cite{evans_20} and it may be relevant in situations in which one seeks to maintain a stochastic process close to a given level.

The paper is organized as follows. In Sect. \ref{sec_general} we summarize the main traits of level-crossing counting along with their application to Gaussian processes. In Sect. \ref{sec_sr} we generalize the renewal equations for resetting to include inertial processes described by two-dimensional random variables $(X(t),\dot X(t))$. In Sect. \ref{sec_crossing} we study how resetting events modify the crossing intensity and apply our results to two relevant examples of physical interest: random acceleration and inertial Brownian motion. For both cases, we show that the stationary crossing intensity displays a maximum as a function of the resetting rate, and we obtain its asymptotic behavior in the limit of large and small resetting rate. We also show that the stationary upcrossing and downcrossing intensities display a similar maximum, and study their asymptotic behaviour as well.
In Sect. \ref{sec_stationary} we obtain the stationary distribution of the complete process with resettings for the two above-mentioned examples. Sect. \ref{sec_mfat} is devoted to obtaining the integral equations for the survival probability of the complete processes, and
a useful approximation for the mean first-arrival time which shows that, under rather general requirements, resettings may reduce such a time. 
 Finally, in Sect.\ref{conclusions} we present our conclusions and in the Appendices we provide details 
of the calculations.

\section{General aspects of the level-crossing intensity}
\label{sec_general}

Consider a one-dimensional random process $X(t)$ and suppose that the velocity $Y(t)=\dot X(t)$ 
exists, and so does
the joint PDF
$p(x,y,t|x_0,y_0,t_0)$ of position and velocity,
\footnote{If not strictly necessary we will omit in what follows the dependence of the initial state $(x_0, y_0)$ in the joint density and other statistics.}
\bd
p(x,y,t|x_0,y_0, t_0)dxdy={\rm Prob}\{ x<X(t)\leq x+dx, y< Y(t)\leq y+dy|X(t_0)=x_0, Y(t_0)=y_0\}\,.
\ed
An important aspect of the level-crossing problem consists in counting the number of crossings of a level $u$ in a interval $[t_0,t]$, which is defined as 
$$
N_u(t,t_0)= \mbox{Number of times} \left[ X(t')=u\right], \quad (t_0\leq t'\leq t).
$$
For some applications it is useful to distinguish between the number of {\it upcrossings}  $N^{(+)}_u(t,t_0)$ (resp. {\it downcrossings} $N^{(-)}_u(t,t_0)$), namely crossings at positive (resp. negative) velocity. Since tangencies to any level $u$ are supposed to be a set of zero measure \cite{lindgren_book}, then the total number of crossings is the sum of upcrossings and downcrossings, 
$$
N_u(t,t_0)=N^{(+)}_u(t,t_0)+N^{(-)}_u(t,t_0).
$$

The crossing intensity is defined as the expected number of crossings per unit time, that is,
\begin{equation}
\mu_u(t)\equiv\lim_{\Delta t\to 0} \frac{\langle N_u(t+\Delta t,t)\rangle}{\Delta t}.
\label{mu_def}
\end{equation}
Similarly, the upcrossing and downcrossing intensities $\mu_u^{(\pm)}(t)$ are defined by replacing $N_u$ with
$N_u^{(\pm)}$ in the above equation.
In Rice's theory one obtains (see e.g. \cite{lindgren_book,maso_pala} for a detailed derivation)  
\begin{equation}
\mu_u(t)=\int_{-\infty}^\infty |y| p(u,y,t) dy,
\label{rice_gen}
\end{equation}
which is the most general expression for the crossing intensity. This expression can be easily modified to
obtain the intensities of upcrossings and downcrossings \cite{maso_pala}:
\begin{equation}
\mu_u^{(\pm)}(t)=\int_{0}^\infty y p(u,\pm y,t) dy,
\label{mu_pm}
\end{equation}
and $\mu_u(t)=\mu_u^{(+)}(t)+\mu_u^{(-)}(t)$. 

If $X(t)$ is a stationary random process, then it is time homogeneous and there exists a nonvanishing and time-independent stationary joint distribution, defined as the limit \cite{maso_llibre}
$$
p_{st}(x,y)=\lim_{t\to\infty}p(x,y,t).
$$
In this case the stationary crossing intensity also exists and is given by
\begin{equation}
\mu_u\equiv \lim_{t\to\infty} \mu_u(t)=\int_{-\infty}^\infty |y| p_{st}(u,y) dy,
\label{mu_stat_def}
\end{equation}
and similarly for the stationary upcrossing and downcrossing intensities $\mu_u^{(\pm)}$ (cf. Eq.~\eqref{mu_pm}). 

Obtaining explicit expressions for the crossing intensity turns out to be quite difficult for non-Gaussian processes \cite{rychlik,lindgren_book} because the joint density $p(x,y,t)$ of position and velocity is usually rather difficult to obtain and, in many cases, it has no known analytical expression. Fortunately this is not the case of Gaussian processes for which the joint PDF is
\begin{equation}
p(x,y,t)=\frac{1}{2\pi\Delta(t)}\exp\biggl\{-\frac{1}{2\Delta^2(t)}\Bigl[\sigma_y^2(t)(x-m_x(t))^2-2\sigma_{xy}(t)(x-m_x(t))(y-m_y(t))+\sigma_y^2(t)(y-m_y(t))^2\Bigr]\biggr\}, 
\label{gauss_pdf}
\end{equation}
where
$$
m_x(t)=\langle X(t)\rangle, \qquad m_y(t)=\langle Y(t)\rangle,
%\label{averages}
$$
and
$$
\sigma_x^2(t)=\left\langle\left[X(t)-m_x(t)\right]^2\right\rangle,\quad  
\sigma_y^2(t)=\left\langle\left[Y(t)-m_y(t)\right]^2\right\rangle, \quad 
\sigma_{xy}(t)=\Bigl\langle [X(t)-m_x(t)][Y(t)-m_y(t)]\Bigr\rangle,
%\label{variances}
$$
are mean values and variances respectively, and
\begin{equation}
\Delta(t)=\sqrt{\sigma_x^2(t)\sigma_y^2(t)-\sigma_{xy}^2(t)}.
\label{Delta_t}
\end{equation}
Note that the dependence on the initial state $(x_0,y_0)$ lies inside the average values $m_x(t)$ and $m_y(t)$.

In the Gaussian case it is possible to get an explicit and general expression for the crossing intensity $\mu_u(t)$ to any level $u$. Indeed, substituting Eq.~\eqref{gauss_pdf} into Eqs.~\eqref{rice_gen} and~\eqref{mu_pm} we obtain \cite{maso_pala}
\begin{equation}
\mu_u(t)=\frac{\Delta(t)}{\pi\sigma_x^2(t)} e^{-(u-m_x(t))^2/2\sigma^2_x(t)}\left\{e^{-\eta_u^2(t)} + \sqrt\pi\eta_u(t){\rm  Erf} \bigl[\eta_u(t)\bigr]\right\},
\label{mu_gauss}
\end{equation}
and 
\begin{equation}
\mu_u^{(\pm)}(t)=\frac{\Delta(t)}{2\pi\sigma_x^2(t)} e^{-(u-m_x(t))^2/2\sigma^2_x(t)}\left\{e^{-\eta_u^2(t)} \pm 
\sqrt\pi\eta_u(t){\rm  Erfc} \bigl[\mp\eta_u(t)\bigr]\right\},
\label{mu_pm__gauss}
\end{equation}
where ${\rm  Erf}(\cdot)$ and ${\rm  Erfc}(\cdot)$ are the error function and the complementary error function respectively, 
and
\begin{equation}
\eta_u(t) \equiv \frac{m_y(t)\sigma_x(t)}{\sqrt 2 \Delta(t)}+\frac{\sigma_{xy}(t)}{\sqrt 2 \Delta(t)\sigma_x(t)}[u-m_x(t)].
\label{eta}
\end{equation} 

In Ref.~\cite{maso_pala} the above formalism was applied to linear inertial processes,
which are described by the following second-order Langevin equation
\begin{equation}
\ddot X(t)+\beta\dot X(t)+\alpha X(t)=k\xi(t),
\label{lang_1}
\end{equation}
where $\alpha$, $\beta$, and $k$ are usually constant parameters (even though they could be functions of time, as in aging processes) and 
$\xi(t)$ is Gaussian white noise with zero mean and unit variance:
$$
\langle\xi(t)\rangle=0, \qquad \langle\xi(t)\xi(t')\rangle=\delta(t-t').
$$
 
Equation~\eqref{lang_1} embraces three cases of significant physical interest: (i) the random acceleration process, where $\alpha=\beta=0$, (ii) the inertial Brownian motion, where $\alpha=0$, $\beta>0$ and (iii) the noisy linear oscillator, where $\alpha > 0, \beta > 0$. Cases (i) and (ii) are nonstationary while case (iii), corresponding to the harmonic oscillator, results in a stationary process. 

Due to the linearity of Eq.~\eqref{lang_1}, as well as the Gaussian character of $\xi(t)$, we see that the process $X(t)$ and its derivative are both Gaussian and their joint PDF, $p(x,y,t)$, is given by Eq.~\eqref{gauss_pdf}. Furthermore, by solving Eq.~\eqref{lang_1} we can obtain explicit expressions of averages and variances which, after substituting for Eqs.~\eqref{mu_gauss} and~\eqref{mu_pm__gauss} allows us to get the crossing intensities $\mu_u(t)$ and 
$\mu_u^{(\pm)}(t)$. Two of us have applied  this procedure in Ref.~\cite{maso_pala} and obtained explicit expressions of the crossing intensities for the cases (i)--(iii) just mentioned, and discussed asymptotic approximations. We refer the reader to Ref. \cite{maso_pala} for details.

\section{Stochastic resetting for inertial processes}
\label{sec_sr}

Let $X(t)$ be an inertial process whose dynamical evolution is governed by a second-order Langevin equation, with initial 
conditions $X(t_0)=x_0$ and $\dot X(t_0)=y_0$, and assume that superimposed to this dynamical evolution there are resetting events which randomly and instantaneously bring $X(t)$ and $Y(t)=\dot X(t)$ to a fixed position and velocity,
$$
(X(t), Y(t)) \quad \longrightarrow \quad (x_r,y_r),
$$
from which the process starts afresh.  Resettings occur at random instants of time.
We suppose that the time intervals between two consecutive events are identically distributed and denote with $\psi(\tau)$ the PDF of one such interval $\tau$.

Let us consider the bidimensional process
$$
\mathbf{Z}(t)=(X(t),Y(t)),
$$
and denote by $p_0(\mathbf z,t|\mathbf z_0,t_0)$ the joint PDF of $(X(t),Y(t))$ 
in the absence of resettings, 
$$
p_0(\mathbf z,t|\mathbf z_0,t_0)dx dy={\rm Prob}\bigl\{ x<X(t)\leq x+dx, y<Y(t)\leq y+dy \ |\ \mathbf Z(t_0)=\mathbf z_0;\ {\rm no}\ {\rm resettings} \bigr\},
$$
where $\mathbf z=(x,y)$, $\mathbf z_0=(x_0,y_0)$, and by $p(\mathbf z,t|\mathbf z_0,t_0)$ the joint PDF of the combined process in the
presence of resettings. In what follows we will assume that the underlying process as well as the resetting mechanism are both time homogeneous, so that
\begin{equation}
p_0(\mathbf z,t|\mathbf z_0,t_0)=p_0(\mathbf z,t-t_0|\mathbf z_0), \qquad {\rm and} \qquad 
p(\mathbf z,t|\mathbf z_0,t_0)=p(\mathbf z,t-t_0|\mathbf z_0).
\label{homo}
\end{equation}
We will see next how $p$ is related to $p_0$. Recall that resettings are instantaneous and assumed to happen on both position $X(t)$ and velocity $Y(t)$. Thus, if at the instant $t'$ the bidimensional process $\mathbf Z$ has reached the value $\mathbf Z(t')=(x',y')$, where 
$$
x'=X(t'-0), \qquad y'=Y(t'-0),
$$
and a resetting occurs, then 
$$
X(t'+0)=x_r, \qquad Y(t'+0)=y_r,
$$
and the bidimensional process starts afresh from $\mathbf z_r=(x_r,y_r)$.  We will assume that resetting events are Poissonian, which implies
$$
\psi(\tau)=r e^{-r\tau},
$$
where $r>0$ is the rate (or frequency) of resetting, so that $r^{-1}$ is the average time interval between two consecutive resettings.  Then, the propagator $p(\mathbf z,t|\mathbf z_0)$ obeys the integral equation \cite{evans_20, mpm_22}
\begin{equation}
p(\mathbf z,t|\mathbf z_0,t_0)=e^{-r(t-t_0)}p_0(\mathbf z,t|\mathbf z_0,t_0)+
r\int_{t_0}^t e^{-r(t-t')}dt' \int_{-\infty}^\infty dx' \int_{-\infty}^\infty p_0(\mathbf z, t|\mathbf z_r,t')p(x',y',t'|\mathbf z_0,t_0) dy',
\label{ie1}
\end{equation}
where the first term on the right-hand side accounts for the evolution with no resetting events between $t_0$ and $t$, while the second term accounts for the probability that the last resetting event was at $t'$ and no reset occurred after $t'$. Let us note that because of the normalization of the joint PDF, i.e., 
$$
\int_{-\infty}^\infty dx' \int_{-\infty}^\infty p(x',y',t'|\mathbf z_0,t_0)dy'=1,
$$
and taking time homogeneity into account (cf. Eq.~\eqref{homo}) we can greatly simplify Eq.~\eqref{ie1} resulting in the expression
\begin{equation}
p(x,y,t | x_0,y_0)=e^{-rt}p_0(x,y,t| x_0, y_0)+
r\int_{0}^t e^{-rt'} p_0(x,y, t'|x_r,y_r)dt',
\label{ie2} 
\end{equation}
which gives the PDF of the complete process with resettings in terms of the PDF of the underlying reset-free process.

\section{Crossing intensity under resettings}
\label{sec_crossing}

We now turn to the problem of determining the crossing intensity in the presence of resettings. Let us denote by $\mu_u^{(0)}(t|x_0,y_0)$ and $\mu_u(t|x_0,y_0)$ the crossing intensities to some level $u$ of the reset-free process and the complete process with resettings, respectively. In terms of their respective joint PDF's 
these intensities are given by Eq.~\eqref{rice_gen},
\begin{equation}
\mu_u^{(0)}(t|x_0,y_0)=\int_{-\infty}^\infty |y| p_0(u,y,t|x_0,y_0) dy, \qquad 
\mu_u  (t|x_0,y_0)=\int_{-\infty}^\infty |y| p(u,y,t|x_0,y_0) dy. 
\label{rice_gen_r}
\end{equation}
For Poissonian resettings, by combining Eq.~\eqref{ie2} with Eq.~\eqref{rice_gen_r}, we obtain
\begin{equation}
\mu_u  (t|x_0,y_0)=e^{-rt}\mu_u^{(0)}(t|x_0,y_0)+r\int_{0}^t e^{-rt'} \mu_u^{(0)}(t'|x_r,y_r)dt',
\label{ie_mu1} 
\end{equation}
which is the most general expression of the crossing intensity under Poissonian resettings in terms of the crossing intensity of the reset-free process.  Proceeding in a similar way we can obtain the upcrossing and downcrossing 
intensities as
$$
\mu_u^{(\pm)}(t|x_0,y_0)=e^{-rt}\mu_u^{(0,\pm)}(t|x_0,y_0)+r\int_{0}^t e^{-rt'} \mu_u^{(0,\pm)}(t'|x_r,y_r)dt',
$$
where $\mu_u^{(0,+)}(t)$ and $\mu_u^{(0,-)}(t)$ denote, respectively, the 
upcrossing and downcrossing intensities of the reset-free process.

In what follows we will assume that resettings bring the process to the initial state, so that $x_r=x_0$ and $y_r=y_0$, and simply denote 
$\mu_u(t)=\mu_u(t|x_0,y_0)$,  and similarly for $\mu_u^{(\pm)}(t)$.\footnote{Let us note that if resettings bring the process to cross the level $u$, such a crossing event is not counted for obtaining $\mu_u(t)$.}
In the $t\to\infty$ limit and assuming that $e^{-rt}\mu_u^{(0)}(t)\to 0$ ($t\to\infty$) we obtain 
the stationary crossing intensity 
\begin{equation}
\mu_u  = r\int_0^\infty e^{-rt}\mu_u^{(0)}(t)dt = r \widehat{\mu}_u^{(0)}(r),
\label{mu_infty}
\end{equation}
in terms of the Laplace transform  $\widehat{\mu}_u^{(0)}(r)$ of the reset-free crossing intensity. Similarly, the stationary upcrossing and downcrossing intensities  are given 
by the Laplace transform of ${\mu}_u^{(0,\pm)}(t)$, multiplied by $r$.

Since resettings usually render the complete process stationary even if the underlying reset-free processes is not, we may have a similar situation for the stationary crossing intensity, namely we may have $\mu_u^{(0)} = 0$ and $\mu_u \neq 0$. For this to happen it suffices that $\mu_u^{(0)}(t)$ behaves appropriately as $t\to 0$ and $t\to\infty$ for the integral in Eq.~\eqref{mu_infty} to exist. 

As shown in Sect.~\ref{sec_general}, when the underlying process is Gaussian it is possible to obtain explicit expressions of the reset-free crossing intensity $\mu_u^{(0)}(t)$, cf. Eq.~\eqref{mu_gauss}. We will now apply the above results to obtain the crossing intensity of two examples of physical interest, the random acceleration process and the inertial Brownian motion when both are under Poissonian resettings.

\subsection{Random acceleration}

Suppose the underlying reset-free inertial process $X(t)$ is given by the second-order Langevin equation:
\begin{equation}
\ddot X(t)=k\xi(t),
\label{leq_ra}
\end{equation}
$k>0$, where $\xi(t)$ is Gaussian white noise and the initial state is $X(0)=x_0, \dot{X}(0)=y_0$. The process is obviously Gaussian and average values and variances are 
\beq
m_x(t)=x_0 + y_0 t, \quad m_y(t) = y_0
\label{av_RA}
\eeq
and
\beq
\sigma_x^2(t) = \frac{k^2 t^3}{3}, \quad
\sigma_y^2(t) = k^2 t, \quad
\sigma_{xy}(t) = \frac{k^2 t}{2}\,.
\label{var_RA}
\eeq
By direct replacement of Eqs.~\eqref{av_RA} and~\eqref{var_RA} into Eq.~\eqref{mu_gauss} one finds  that the reset-free crossing intensity to any level $u$ is given by~\cite{maso_pala}
\begin{equation}
\mu_u^{(0)}(t)=\frac{\sqrt 3}{2\pi t} e^{-3(u-m_x(t))^2/2k^2t^3}\left[e^{-\eta_u^2(t)}+ \sqrt\pi \eta_u(t) {\rm Erf} (\eta_u(t))\right], 
\label{mu_ra_0}
\end{equation}
where, cf. Eq.~\eqref{eta}, 
\begin{equation}
\eta_u(t)=\frac{\sqrt 2}{kt^{1/2}}\left[y_0+\frac{3}{2t}(u-m_x(t))\right].
\label{eta_ra}
\end{equation}

The asymptotic expression as $t\to\infty$ to any level $u$ and for any value of the initial velocity $y_0$ is given by
\begin{equation}
\mu^{(0)}_u(t) \simeq\frac{\sqrt 3}{2\pi t}, \qquad (t\to\infty).
\label{mu_ra_asym}
\end{equation}
We therefore see that the stationary 
crossing intensity of the reset-free random acceleration process is zero.

For the complete process with Poissonian resettings, the crossing intensity $\mu_u(t)$ is finite provided $u$ and $y_0$ do not both vanish, and
we show below that in this case it has a nonzero stationary value.  On the other hand, if $u=y_0=0$, the reset-free crossing intensity
in Eq.~\eqref{mu_ra_0} becomes $\mu_{u=0}^{(0)}(t)=\sqrt{3}/2 \pi t$ at all times,
and substituting this into Eq.~\eqref{ie_mu1} we obtain an integral that diverges at short times, thus $\mu_{u=0}(t)=\infty$, which reflects the fact that after each resetting there is a high probability of crossing the zero level.

\subsubsection{Case $y_0 = 0$, $u\neq 0$}

\begin{figure}[ht]
\includegraphics[width=0.6\linewidth,angle=0]{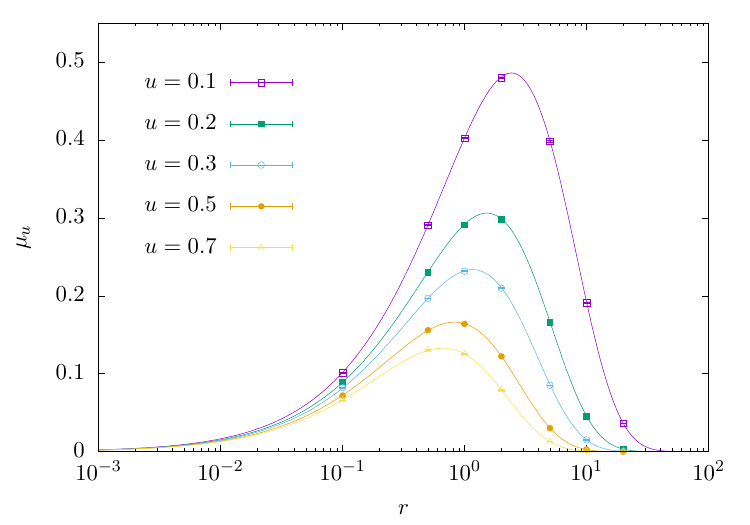}
\caption{(Color online)  Stationary crossing intensity $\mu_u$ for the random acceleration process, with $y_0=0, k=1$ and for different values of $u$, as a function of the resetting intensity $r$. The lines are obtained by numerically integrating Eq.~\eqref{mu_RA_y0_reset}. The points are the results of Monte Carlo simulation of the Langevin equation.}
\label{fig_mu_RA_y0}
\end{figure}

We will focus first on the case of zero initial velocity, $y_0=0$, and nonzero level $u\neq 0$. In this case two timescales appear in the problem: (i) the average time $r^{-1}$ between two consecutive resettings and (ii) 
\beq
\tau_u \equiv \left(\frac{|u|}{k}\right)^{2/3} \, ,
\label{tau_u}
\eeq
which is the timescale at which the process reaches level $u$, namely $X(\tau_u)\simeq \pm u$, starting from $x_0=0$.
Substituting Eq.~\eqref{mu_ra_0} (with $y_0=m_x(t)=0$) into Eq.~\eqref{mu_infty} we obtain the stationary crossing intensity
\beq
\mu_u  = \frac{\sqrt{3}}{2 \pi}r \int_0^\infty \frac{1}{t}\left[e^{-\frac{9}{2} (\tau_u/t)^3}+ 3  \sqrt{\frac{\pi}{2}}  \left(\frac{\tau_u}{t}\right)^{3/2}
\mbox{Erf}\left(\frac{3}{\sqrt{2}}(\tau_u/t)^{3/2}\right)\right] e^{-r t -\frac{3}{2} (\tau_u/t)^3} dt, \qquad (y_0=0)\,.
\label{mu_RA_y0_reset}
\eeq 
Figure \ref{fig_mu_RA_y0} shows that $\mu_u$, obtained by the numerical evaluation of the above integral, displays a maximum as a function of $r$. The existence of this maximum can be understood by noting that as $r$ tends to zero the process wanders further and further from any fixed level, making a crossing increasingly less probable, while for $r$ large and growing it becomes increasingly more likely to incur in a resetting before it can cross a level $u$. Notice also that $\mu_u$ increases as $u$ decreases, and diverges logarithmically as $u\to 0$ for all $r$. 

In order to check our results, we also performed  Monte Carlo simulations of the Langevin equation Eq.~\eqref{leq_ra}, by adapting the algorithm of Ref.~\cite{farago}, as discussed in Ref.~\cite{maso_pala}. As shown in Fig.~\ref{fig_mu_RA_y0}, the numerical integration and the Monte Carlo data agree perfectly.
Here and in the rest of the figures of this paper, the error bars of the Monte Carlo data are smaller than the symbols, all timescales are expressed in arbitrary units of time,
and all crossing intensities are expressed in the inverse of the arbitrary units of time.

\begin{figure}[ht]
\includegraphics[width=0.6\linewidth,angle=0]{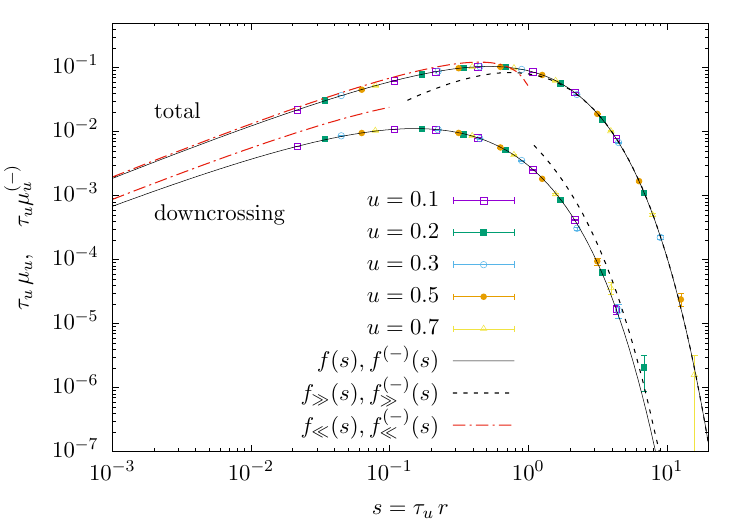}
\caption{(Color online) Scaling and asymptotics of the stationary crossing intensity 
for the random acceleration process, with $y_0=0, k=1$.
Upper curve: the symbols represent the same Monte Carlo data of of Fig.~\ref{fig_mu_RA_y0}, rescaled according to the scaling form in Eq.~\eqref{scaling_RA_y0}. The solid (black) line is obtained by numerical integration of $f(s)$ in Eq.~\eqref{scaling_RA_y0}. The (black) dashed line represents the large-$s$ asymptotic expression for the total intensity, given in Eq.~\eqref{large_r_RA_y0}. 
The (red) dot-dashed line represents the small-$s$ asymptotic expression for the total intensity, given in Eq.~\eqref{small_r_RA_y0}.
Lower curve: the symbols represent the Monte Carlo results for the 
stationary downcrossing intensity, for the same values of $u$ as in the upper curve, 
rescaled according to Eq.~\eqref{mupm_RA_y0}. The solid (black) line is obtained
by numerical integration of $f^{(-)}(s)$ in Eq.~\eqref{mupm_RA_y0}.
The (black) dashed line represents the large-$s$ asymptotic expression for the downcrossing intensity, given in Eq.~\eqref{large_mupm_RA_y0}. 
The (red) dot-dashed line represents the small-$s$ asymptotic expression for the downcrossing intensity, given in Eq.~\eqref{small_mupm_RA_y0}.
}
\label{fig_mu_RA_y0_scaling}
\end{figure}

The curves in Fig.~\ref{fig_mu_RA_y0} for different values of $u$ can be rescaled onto a single curve according to the scaling form 
\beq
\mu_u = \frac{1}{\tau_u}f(\tau_u r), \qquad 
f(s) \equiv \frac{s \sqrt{3}}{2 \pi} \int_0^\infty \frac{1}{\theta}\left[e^{-\frac{9}{2 \theta^3}}+ 3 \, \sqrt{\frac{\pi}{2\theta^3}} 
\mbox{Erf}\left(\frac{3}{\sqrt{2 \theta^3}}\right)\right] e^{-s \theta -\frac{3}{2 \theta^3}} d\theta ,
\label{scaling_RA_y0}
\eeq 
as shown in Fig.~\ref{fig_mu_RA_y0_scaling}.

The above scaling relation implies that the position 
of the maximum scales as the inverse of $\tau_u$. The proportionality constant can be obtained in principle by the integral equation that arises from imposing $f^\prime(s)=0$, but we have not attempted this.  We have instead obtained the asymptotic behavior of $f(s)$ for both 
large $s$ and small $s$.
For large $s$,
evaluating the integral in Eq.~\eqref{scaling_RA_y0} with the Laplace method~\cite{erderlyi} 
(see Appendix \ref{app_f} for details)  gives the asymptotically exact result

\beq
f(s) \underset{s \to \infty}{\simeq} f_{\gg}(s) \equiv \frac{s}{2} \sqrt{\frac{3}{2}}\mbox{Exp}\left(-\frac{2^{7/4}}{\sqrt{3}} s^{3/4}\right)
\Erf\left(\frac{3^{1/4}}{2^{1/8}} s^{3/8}\right),
\label{large_r_RA_y0}
\eeq
which, as shown in Fig.~\ref{fig_mu_RA_y0_scaling}, approximates very well the numerically integrated $f(s)$ for large $s$, with a relative deviation of less 
than 1\% already for $s \simeq 10$. 
Obtaining an exact asymptotic expression for small $s$ is more involved. In this case, in Appendix \ref{app_f} we obtain the following approximation
\beq
f(s)\underset{s\to 0}\simeq f_{\ll}(s) \equiv s \frac{\sqrt{3}}{2 \pi} 
 \left[e^{-1/3} \sqrt{\frac{\pi}{6}} - \gamma - \ln\left(s \, 18^{1/3}\right)\right] + \frac{s}{3},
\label{small_r_RA_y0}
\eeq
where $\gamma$ is the Euler-Mascheroni constant. This agrees qualitatively with the numerically integrated $f(s)$, as shown in Fig.~\ref{fig_mu_RA_y0_scaling}, 
with a relative deviation smaller than 20\% in the range $s\in [10^{-3}, 2 \times 10^{-1}]$.

It is interesting to study also the stationary upcrossing and downcrossing intensities. Using Eqs.~\eqref{mu_pm__gauss} and~\eqref{mu_ra_0}, and
proceeding as above we obtain 
\beq
\mu_u^{(\pm)} = \frac{1}{\tau_u}f^{(\pm)}(\tau_u r), \qquad
f^{(\pm)}(s)\equiv 
\frac{s \sqrt{3}}{4 \pi}
 \int_0^\infty \frac{1}{\theta}\left[e^{-\frac{9}{2 \theta^3}} \pm 3 \,\sgn(u) \sqrt{\frac{\pi}{2\theta^3}} 
\mbox{Erfc}\left(\mp \frac{3\,\sgn(u)}{\sqrt{2 \theta^3}}\right)\right] e^{-s \theta -\frac{3}{2 \theta^3}} d\theta ,
\label{mupm_RA_y0}
\eeq
where $\sgn(u)$ is the sign function ($\sgn(u)=1$ if $u>0$ and $\sgn(u)=-1$ if $u<0$). Using the identity $\mp \mbox{Erfc}(\pm z)=\mbox{Erf}(z)\mp 1$, we obtain the following relation between the directional and total stationary crossing intensities:
\beq
\mu_u^{(\pm)} = \frac{\mu_u}{2}\pm \sgn(u) \frac{3 r}{4} \sqrt{\frac{3}{2 \pi}} 
\int_0^\infty \frac{1}{t} \left(\frac{\tau_u}{t}\right)^{3/2}
e^{-r t -\frac{3}{2} (\tau_u/t)^3} dt \,.
\label{mupm_correction}
\eeq
We note that for $u>0$ we have $\mu_u^{(+)} > \mu_u^{(-)}$ because, even if the process has zero mean, after an upcrossing a resetting can occur before a downcrossing can take place. In other words, a downcrossing is always preceded by an upcrossing, but the reverse is not necessarily true. This effect becomes more important as the 
resetting rate increases: indeed, by evaluating $f^{(+)}(s)$ with the Laplace method in a manner similar to that used for $f(s)$ in Appendix \ref{app_f}, we obtain that 
$f^{(+)}(s)/f(s) \to 1$ as $s \to \infty$, namely all crossings are upcrossings for large $s$. Note also that by symmetry, for $u<0$ the values of the upcrossing and downcrossing intensities are reversed, i.e. $\mu_u^{(+)} = \mu_{-u}^{(-)}$.

The stationary downcrossing intensity for $u>0$, obtained by numerically integrating $f^{(-)}(s)$ in Eq.~\eqref{mupm_RA_y0}, is shown in Fig.~\ref{fig_mu_RA_y0_scaling} and agrees perfectly with Monte Carlo simulations. Using the Laplace method (see Appendix \ref{app_fm} for details) we obtain
\beq
f^{(-)}(s) \underset{s \to \infty}{\simeq} f^{(-)}_{\gg}(s) \equiv \frac{1}{\sqrt{\pi} (2^{15} 9 s)^{1/8}}\mbox{Exp}\left(-\frac{2^{9/4}}{\sqrt{3}} s^{3/4} \right),
\label{large_mupm_RA_y0}
\eeq
which gives the correct qualitative behavior, as shown in Fig.~\ref{fig_mu_RA_y0_scaling}, albeit with a relative deviation around 50\%  in the range of $s$ displayed in the figure. Finally, for small $s$, using again the Laplace method, in Appendix \ref{app_fm} we obtain the approximate expression
\beq
f^{(-)}(s) \underset{s \to 0}{\simeq} f^{(-)}_{\ll}(s) \equiv
s \frac{\sqrt{3}}{4 \pi} \left[ e^{-1/3} \sqrt{\frac{\pi}{6}} - \gamma - \frac{1}{3}\ln \frac{9}{2}-\ln s\right],
\label{small_mupm_RA_y0}
\eeq
which reproduces the correct qualitative behavior, with a relative deviation between 20\% and 50\% in the range of $s$ displayed in the figure.

\subsubsection{Case $y_0 \neq 0$, $u = 0$}

\begin{figure}[ht]
\includegraphics[width=0.6\linewidth,angle=0]{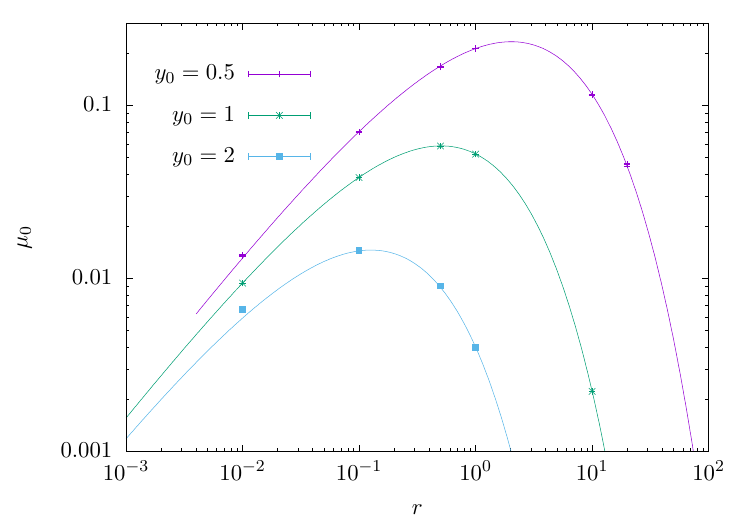}
\caption{(Color online) Stationary zero-crossing intensity $\mu_0$ for the random acceleration process with 
$k=1$ and different initial velocities $y_0$,
 as a function of the resetting intensity $r$. The lines are obtained by numerically integrating Eq.~\eqref{mu_ra1}. The points are the results of Monte Carlo simulation of the Langevin equation.}
\label{fig_RA_u0_y}
\end{figure}

\begin{figure}[ht]
\includegraphics[width=0.6\linewidth,angle=0]{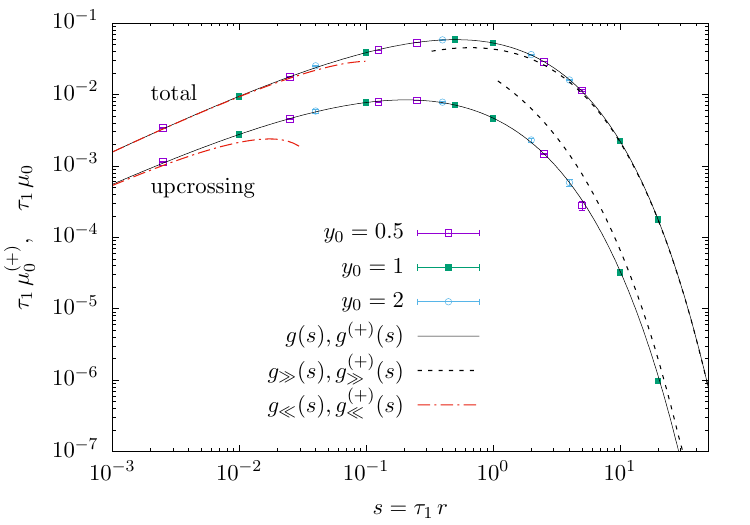}
\caption{(Color online) Scaling and asymptotics of the stationary zero-crossing intensity 
for the random acceleration process, with $u=0, k=1$.
Upper curve: the symbols represent the same Monte Carlo data of of Fig.~\ref{fig_RA_u0_y}, rescaled according to the scaling form in Eq.~\eqref{mu_ra1}. The solid (black) line is obtained by numerical integration of $g(s)$ in Eq.~\eqref{mu_ra1}. The (black) dashed line represents the large-$s$ asymptotic expression for the total intensity, given in Eq.~\eqref{mu_ra_large_main}. The (red) dot-dashed line represents the small-$s$ asymptotic expression for the total intensity, given in Eq.~\eqref{mu_ra_small2_main2}.
Lower curve: the symbols represent the Monte Carlo results for the stationary upcrossing intensity, for the same values of $y_0$ as in the upper curve, 
rescaled according to Eq.~\eqref{mupm_RA_u0_y}. The solid (black) line is obtained by numerical integration of $g^{(+)}(s)$ in Eq.~\eqref{mupm_RA_u0_y}.
The (black) dashed line represents the large-$s$ asymptotic expression for the upcrossing intensity, given in Eq.~\eqref{large_up_u0}. 
The (red) dot-dashed line represents the small-$s$ asymptotic expression for the upcrossing intensity, given in Eq.~\eqref{small_up_u0}.}
\label{fig_RA_u0_y_scaling}
\end{figure}

We now turn to the case of non-zero initial velocity, $y_0\neq 0$, and consider the crossing at level $u=0$. In this case the timescales of the problem are $r^{-1}$ and
\beq
\tau_1 \equiv \frac{y_0^2}{k^2},
\label{tau_1}
\eeq
which is the time at which the standard deviation of the velocity becomes of the order of the initial velocity, $\sigma_y(\tau_1)\simeq |y_0|$.
As in the previous case, the stationary zero-crossing intensity $\mu_0$ is obtained by substituting Eq.~\eqref{mu_ra_0} (with $u=0, m_x(t)=y_0 t$) into Eq.~\eqref{mu_infty}, which in the limit $t\to \infty$ yields
\begin{equation}
\mu_0 =\frac{1}{\tau_1} g(\tau_1 r), \qquad
g(s) \equiv \frac{s \sqrt{3}}{2\pi}\int_0^\infty \frac{1}{\theta}\left[e^{-1/\theta}+ 
 \sqrt{\frac{\pi}{\theta}} {\rm Erf} \left(1/\sqrt{\theta}\right) \right]e^{-3/\theta-s  \theta/2} d\theta,
\label{mu_ra1}
\end{equation}
where we have made the change of variables $\theta = 2 t /\tau_1$ in the integral. Figure \ref{fig_RA_u0_y} shows $\mu_0$, obtained by numerically integrating $g(s)$, as a function of $r$, tested against Monte Carlo simulations, with perfect agreement as otherwise expected. We observe again a maximum as a function of $r$, which can be understood in the same way we have discussed above for the previous case $y_0=0, u\neq 0$. Note also that $\mu_0$ increases as $y_0$ decreases and eventually diverges logarithmically for $y_0 \to 0$. In Fig.~\ref{fig_RA_u0_y_scaling} we show the data rescaled according to Eq.~\eqref{mu_ra1}.

In Appendix~\ref{app_g_exact} we prove that the exact analytical expression for the integral in Eq.~\eqref{mu_ra1} can be written in terms of an infinite sum of modified Bessel functions of the second kind~\cite{mos}, $K_n(z)$, 
\begin{equation}
g(s)=\frac{s \sqrt{3}}{\pi} \left[ K_0\left(2\sqrt{2 s }\right)+ 2 \sum_{n=0}^{\infty} \frac{(-1)^n}{n!}\frac{1}{2n +1} \left(\frac{s }{6}\right)^{\frac{n+1}{2}}  K_{n+1}\left(\sqrt{6 s }\right) \right],
\label{mu_ra_exact_main}
\end{equation}
and how this formula can be readily used to obtain asymptotic approximations valid for large and small values of $s$. Nevertheless, these approximate expressions can be also recovered by direct inspection of Eq.~\eqref{mu_ra1}, as we will show. 

Thus, for large $s$, the integral in Eq.~\eqref{mu_ra1} is dominated by small values of $\theta$, at which 
the first term inside the square brackets is negligible compared to the second. Therefore, since 
${\mbox{Erf}}(x)\sim x$ for $x\to 0$, we have
\beq
g(s) \underset{s \to \infty}{\simeq} g_{\gg}(s)\equiv  \frac{s}{2}\sqrt{\frac{3}{\pi}}\int_{0}^\infty 
\frac{1}{\theta^{3/2}} e^{-3/\theta - s \theta/2} d\theta 
=  \frac{s}{2}   e^{-\sqrt{6 s }}\,.
\label{mu_ra_large_main}
\eeq
where the last integral was evaluated exactly~\cite{roberts}.
Conversely, for small $s$ the integral in Eq.\eqref{mu_ra1} is dominated by large values of $\theta$, 
at which the second term inside the square brackets is negligible, giving
\bd
g(s) \underset{s \to 0}{\simeq} \frac{s}{2}\sqrt{\frac{3}{\pi}}\int_{0}^\infty \frac{1}{\theta} e^{-4/\theta - s \theta/2} d\theta = \frac{s \sqrt{3}}{\pi} K_0(2 \sqrt{2 s}). 
\ed
Since $K_0(x)\sim -2 \gamma - \ln (x/2)$ as $x\to 0$ \cite{roberts}, we have
\bd
g(s) \underset{s \to 0}{\simeq} \frac{s \sqrt{3}}{\pi}\left[- \frac{1}{2}\ln(2 s )-\gamma\right]\, .
\ed
By expanding the error function in Eq.~\eqref{mu_ra1} in power series, as shown in Appendix \ref{app_g_exact}, we obtain the following more accurate expression:
\begin{equation}
g(s) \underset{s \to 0}{\simeq} g_{\ll}(s) \equiv \frac{s \sqrt{3}}{\pi}\left[- \frac{1}{2}\ln(2 s )-\gamma+\frac{\pi}{6\sqrt{3}}\right]\, .
\label{mu_ra_small2_main2} 
\end{equation}
Both expressions in Eq.\eqref{mu_ra_large_main}
and~\eqref{mu_ra_small2_main2} are asymptotically exact, and agree very well with the numerically integrated $g(s)$, as shown in Fig.~\ref{fig_RA_u0_y_scaling}, with a relative deviation below $1\%$ for $s \gtrsim 20$ and $s\lesssim 0.006$ for $g_\gg$ and $g_\ll$ respectively.

Proceeding as in the case $y_0=0, u\neq 0$ we also obtain the stationary zero-upcrossing and zero-downcrossing intensities:
\beq
\mu_0^{(\pm)} =\frac{1}{\tau_1} g^{(\pm)}(\tau_1 r), \qquad
g^{(\pm)}(s) \equiv \frac{s \sqrt{3}}{4\pi}\int_0^\infty \frac{1}{\theta}\left[e^{-1/\theta}\mp
\sgn(y_0) \sqrt{\frac{\pi}{\theta}} {\rm Erfc} \left(\pm\sgn(y_0) /\sqrt{\theta}\right) \right]e^{-3/\theta-s \theta/2} d\theta.
\label{mupm_RA_u0_y}
\eeq

Substituting $\mbox{Erfc}(\pm z)=1\mp \mbox{Erf}(z)$ in the above equation and comparing it with Eqs.~\eqref{mu_ra1} and~\eqref{mu_ra_large_main}
we obtain the following relation, valid for all $s$:
\beq
g^{(\pm)}(s)=\frac{1}{2} g(s) \mp \sgn(y_0) \frac{s}{4} e^{-\sqrt{6 s}}.
\label{exact_g_relation}
\eeq
We note that for $y_0>0$ we have $\mu^{-}_0 > \mu^{+}_0$, which can be understood observing that the first zero-crossing after a resetting is necessarily a downcrossing, 
since the process starts at positive velocity. From Eqs.~\eqref{exact_g_relation} and~\eqref{mu_ra_large_main} we see that $g^{(-)}(s)/g(s) \to 1$ as $s \to \infty$, since for large 
reset rate, after the first downcrossing the process does not have enough time do an upcrossing before the next resetting takes place. We also note that,  due to symmetry, the values of $\mu^{(+)}_0$ and $\mu^{(-)}_0$ are swapped when we change the sign of $y_0$. The bottom curve in Fig.~\ref{fig_RA_u0_y_scaling} shows the upcrossing intensity for different values of $y_0$, rescaled according to Eq.~\eqref{mupm_RA_u0_y} where $g^{(+)}(s)$ was integrated numerically, together with the Monte Carlo data.

For the upcrossing intensity at $y_0>0$ we obtain in Appendix \ref{app_ra_u0} the following asymptotic relations:
\beq
g^{(+)}(s)  \underset{s \to \infty }{\simeq} g_{\gg}(s) \equiv 
\sqrt{\frac{\pi}{3}} \left(\frac{s}{2^7}\right)^{1/4}   e^{-2\sqrt{2 s}},
\label{large_up_u0}
\eeq
and
\beq 
g^{(+)}(s) \underset{s \to 0}{\simeq} g_{\ll}(s) \equiv 
\frac{s \sqrt{3}}{2 \pi}\left[- \frac{1}{2}\ln(2 s )-\gamma-\frac{\pi}{3\sqrt{3}}\right]\, ,
\label{small_up_u0}
\eeq
where the latter follows directly from Eqs.~\eqref{mu_ra_small2_main2} and~\eqref{exact_g_relation}. Both relations are displayed in Fig.~\ref{fig_RA_u0_y_scaling} and agree fairly well with the numerically integrated $g(s)$.

Finally, we note that in the general case where $y_0$ and $u$ are nonzero, both timescales, $\tau_u$ and $\tau_1$, participate in the expressions of the crossing intensity and a more complex scaling form is obtained.

\subsection{Inertial Brownian motion}

As a second example we consider as the underlying reset-free process the inertial Brownian motion, described by the Langevin equation
\beq
\ddot X(t)+\beta X(t)=k\xi(t),
\label{LangevBM}
\eeq
where $\beta>0$ is the damping constant, $\xi(t)$ is zero-mean Gaussian white noise, and the initial state is $X(0)=x_0,\dot{X}(0)=y_0$. 
Average values and variances are \cite{maso_pala}: 
$$
m_x(t)=x_0+\frac{y_0}{\beta}\left(1-e^{-\beta t}\right), \qquad m_y(t)=y_0e^{-\beta t},
$$
and
$$
\sigma_x^2(t)=\frac{k^2}{\beta^3}\left(\beta t-\frac 32 + 2e^{-\beta t}-\frac 12 e^{-2\beta t}\right),\quad
\sigma_y^2(t)=\frac{k^2}{2\beta}\left(1-e^{-2\beta t}\right),\quad
\sigma_{xy}(t)=\frac{k^2}{\beta^2}\left(\frac 12 - e^{-\beta t}+\frac 12 e^{-2\beta t}\right).
$$
The exact expression for the reset-free crossing intensity $\mu^{(0)}_u(t)$ is obtained after substituting these expressions into Eq.~\eqref{mu_gauss}, along with the expressions for $\Delta(t)$ and $\eta_u(t)$ given by Eqs.~\eqref{Delta_t} and~\eqref{eta} respectively. All of this results in a rather cumbersome expression which we will not write.  We will obtain instead approximate expressions valid for large and small times. Specifically, for $\beta t\gg 1$, to leading order we have 
\begin{equation}
m_x(t) \simeq x_0+\frac{y_0}{\beta}, \qquad m_y(t) \simeq 0,
\label{asym_m}
\end{equation}
and
\begin{equation}
\sigma_x^2(t) \simeq \frac{k^2 t}{\beta^2},
 \qquad \sigma_y^2(t) \simeq \frac{k^2}{2\beta}, \qquad \sigma_{xy}(t) \simeq \frac{ k^2}{2\beta^2},
\label{asym_sigma}
\end{equation} 
from which we obtain 
\beq                          
\Delta(t)\simeq \frac{k^2 t^{1/2}}{\sqrt 2 \beta^{3/2}}, \qquad 
\frac{\Delta(t)}{\sigma_x^2(t)}\simeq\left(\frac{\beta}{2t}\right)^{1/2}.
\label{asym_delta}
\eeq
In what follows we will consider for simplicity only the case $x_0 = y_0 = 0$. In this case, for large values of $t$ we have
$$
\eta_u(t)\simeq \frac{\beta^{1/2} u}{2 k t} \ , \qquad\quad (\beta t \gg 1).
$$
In the opposite limit $\beta t \ll 1$, the variances tend to the expressions given for the random acceleration process, Eq.~\eqref{var_RA}, and so does $\eta_u(t)$ after setting $x_0 = y_0 = 0$ in Eq.~\eqref{eta_ra}, namely
$$
\eta_u(t) \simeq \frac{3 u}{\sqrt{2} k t^{3/2}}\ , \qquad\quad (\beta t \ll 1).
$$
Therefore in this limit the reset-free crossing intensity $\mu_u^{(0)}(t)$ takes the same form as that of random acceleration (cf. Eq.~\eqref{mu_ra_0} after setting $m_x(t)=0$), a fact that can be understood physically by noting that damping is negligible in the ballistic regime $\beta t \ll 1$.

\begin{figure}[ht]
\includegraphics[width=0.6\linewidth,angle=0]{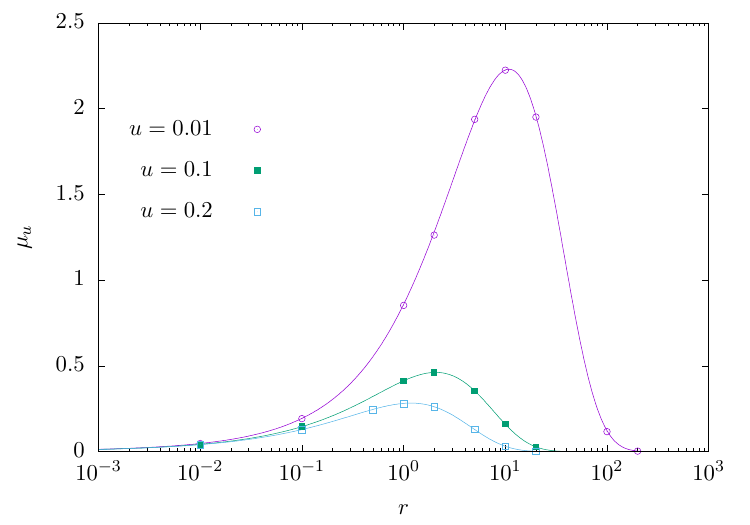}
\caption{(Color online)  Stationary crossing intensity $\mu_u$ for the inertial Brownian motion, with $x_0=y_0=0$, $k=\beta=1$ and for different values of $u$, as a function of the resetting intensity $r$. The lines are obtained by numerical integration of Eq.~\eqref{mu_infty} using the full expression of the reset-free crossing intensity. The points are the results of Monte Carlo simulation of the Langevin equation, Eq.~\eqref{LangevBM}.}
\label{fig_mu_BM_y0}
\end{figure}

For the inertial Brownian motion with Poissonian resettings, we evaluate $\mu_u$ as a function of $r$  by inserting the full expression of $\mu_u^{(0)}(t)$ into 
Eq.~\eqref{mu_infty} and computing the integral in $t$ numerically. The results of the numerical integration, shown in Fig.~\ref{fig_mu_BM_y0} together with our Monte Carlo estimates, display a maximum which increases for decreasing $u$ and diverges as $u\to 0$, a behavior which can be understood in the same way discussed earlier for the random acceleration process.

We note that, compared to random acceleration, an additional timescale $\beta^{-1}$ appears in Brownian motion, which is the time when the damping force becomes of the order of the inertial force. As a result, the stationary crossing intensity with $y_0=0$ will obey a more complicated scaling form
$$
\mu_u = \frac{1}{\tau_u} h(\tau_u r, \tau_u \beta)\,,
$$
and the data of Fig.~\ref{fig_mu_BM_y0} cannot be rescaled on a single curve. 
In the case $y_0 \neq 0$, which we will not consider here, one must consider also the timescale $\tau_1$
\cite{maso_pala}, as in random acceleration.

\begin{figure}[ht]
\includegraphics[width=0.6\linewidth,angle=0]{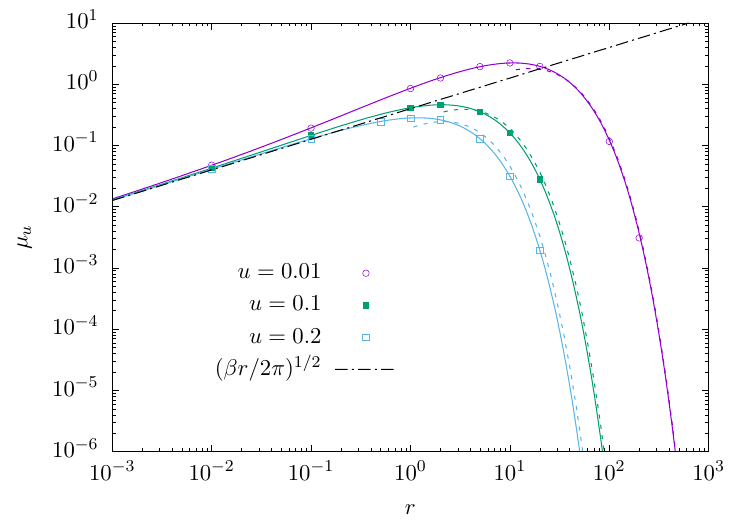}
\caption{(Color online) Asymptotic behavior of the stationary crossing intensity for the inertial Brownian motion.
The symbols and the solid lines are the same as in Fig.~\ref{fig_mu_BM_y0}, displayed in log-log form. The (colored) dashed lines represent the large-$r$ asymptotic expression in Eq.~\eqref{large_BM}. The (black) dot-dashed line represents the small-$r$ asymptotic expression in Eq.~\eqref{small_BM}.}
\label{fig_mu_BM_asympt}.
\end{figure}

Obtaining the large-$r$ asymptotic behavior of $\mu_u$ for arbitrary values of $u$ 
and $\beta^{-1}$ is a complicated task. We obtain an approximation for  this behavior by observing that for large $r/\beta$, the integral in Eq.~\eqref{mu_infty} is dominated by small times.
We therefore replace the full expression of $\mu_u^{(0)}$ with its asymptotic form for small times, given by Eq.~\eqref{mu_ra_0} with $m_x=y_0=0$, and obtain in this way the same expression 
as in random acceleration for the case
$y_0=0, u\neq 0$, given by Eq.~\eqref{mu_RA_y0_reset}.
We can then use the large-$r$ limit of that expression, given in Eq.~\eqref{large_r_RA_y0}, which gives
\beq
\mu_u \underset{r/\beta \gg 1}{\simeq} \frac{r}{2} \sqrt{\frac{3}{2}} \mbox{Exp}\left(-\frac{2^{7/4}}{\sqrt{3}} (\tau_u r)^{3/4}\right)
\Erf\left(\frac{3^{1/4}}{2^{1/8}} (\tau_u r)^{3/8}\right).
\label{large_BM}
\eeq
This approximation becomes asymptotically exact for $\tau_u \beta \to 0$ and $\tau_u r \to \infty$, as shown in Fig.~\ref{fig_mu_BM_asympt}, since in this regime we have $r^{-1}\ll \tau_u \ll \beta^{-1}$ thus damping is negligible.

In the opposite limit $r/\beta \ll 1$, the integral in Eq.~\eqref{mu_infty} will be dominated by large times, and thus we can replace the reset-free intensity with its large $\beta t$ form, 
$$
\mu_u^{(0)}(t) \underset{\beta t\gg 1}{\simeq}  \frac{1}{\pi}\left(\frac{\beta}{2 t}\right)^{1/2}
{\mbox{Exp}}\left(-\frac{u^2 \beta^2}{2 k^2 t} \right)
\left[{\mbox{Exp}}\left(-\frac{u^2 \beta}{2 k^2 t^2} \right)
+ \sqrt{\pi} \frac{u \beta^{1/2}}{\sqrt{2} k t}
{\mbox{Erf}}\left(\frac{u \beta^{1/2}}{\sqrt{2} k t}\right)\right].
$$
Furthermore, the second term can be neglected for large times, hence we obtain 
$$
\mu_u \underset{r/\beta \ll 1}{\simeq} \left(\frac{\beta}{2}\right)^{1/2}\frac{r}{\pi}\int_0^\infty dt \frac{e^{-r t}}{t^{1/2}}
{\mbox{Exp}}\left( -\frac{u^2 \beta^2}{2 k^2 t} -\frac{u^2 \beta}{2 k^2 t^2} \right)
$$
To leading order, the above integral can be evaluated asymptotically as
\beq
\mu_u \underset{r/\beta \ll 1}{\simeq} 
\left(\frac{\beta}{2}\right)^{1/2}\frac{r}{\pi}\int_0^\infty dt \frac{e^{-r t}}{t^{1/2}}
= \left(\frac{\beta r}{2 \pi}\right)^{1/2}\,.
\label{small_BM}
\eeq
This form agrees quite well with the numerically integrated $\mu_u$ for small $r$, as shown in Fig.~\ref{fig_mu_BM_asympt}. Finally, we note that the stationary upcrossing and downcrossing intensities can be evaluated in a manner similar to the one we followed for the random acceleration process.

\section{The stationary distribution}
\label{sec_stationary}

We will now set aside the level crossing counting and turn our attention to the classical issues of resettings, namely, the existence of stationary distributions and the mean-first arrival time. 

Let us first address the problem of determining the stationary distribution in the presence of
resetting. This is defined by the limit 
$$
p_{st}(x,y)=\lim_{t\to\infty} p(x,y,t|x_0,y_0),
$$
as long as it is finite and nonzero. From Eq.~\eqref{ie2} we therefore obtain
\begin{equation}
p_{st}(x,y)=r\int_0^\infty e^{-rt} p_0(x,y,t|x_r,y_r)dt = r\hat p_0(x,y,r|x_r,y_r),
\label{p_stat}
\end{equation}
where $\hat p_0(x,y,r|x_r,y_r)$ is the Laplace transform of the reset-free joint PDF. We observe that the stationary distribution of the process under resettings, 
$p_{st}(x,y)$, may exist even though the reset-free process has no stationary distribution, provided the above Laplace transform exists, a less restrictive condition than the existence of a stationary distribution for the reset-free process  \cite{majumdar_13jpa,maso_mont_2019}. 

Suppose that the reset-free process $X(t)$ is Gaussian. In this case the joint PDF $p(x,y,t|x_0,y_0)$ is explicitly given by Eq.~\eqref{gauss_pdf} and the marginal distribution of $X(t)$ reads
\begin{equation}
p_0(x,t|x_r,y_r)=\frac{1}{\sqrt{2\pi}\sigma_x(t)}\exp\left\{-\frac{[x-m_x(t|x_r,y_r)]^2}{2\sigma_x^2(t)}\right\}.
\label{gauss_p_x}
\end{equation}

Note that the stationary PDF of the reset-free process, defined as the limit   
$$
p_{st}^{(0)}(x)=\lim_{t\to\infty}p_0(x,t|x_0,y_0),
$$
will exist (and it is nonzero) depending on the asymptotic behavior of the variance $\sigma_x(t)$ as $t\to\infty$.

The stationary distribution for the resetting process will thus be given by
\begin{equation}
p_{st}(x|x_r,y_r)=\frac{r}{\sqrt{2\pi}}\int_0^\infty \frac{dt}{\sigma_x(t)}\exp\left\{-rt-\frac{[x-m_x(t|x_r,y_r)]^2}{2\sigma_x^2(t)}\right\}.
\label{gauss_p_st}
\end{equation}
We remark that this density may exist (and it is nonzero) even if $p_{st}^{(0)}(x)$ does not. Below we will illustrate this with the two examples considered in Section IV, random acceleration and inertial brownian motion. In both cases we assume  Poissonian resettings to a fixed state $(x_r,y_r)$, and for simplicity we take $y_r=y_0=0$, so that the process starts at rest and resettings bring it particle to position $x_r$ with zero velocity.

\subsection{Random acceleration}

For the random acceleration process, we have shown elsewhere \cite{maso_pala} that 
the reset-free distribution is
\beq
p_{0}(x,t|x_r)=\sqrt{\frac{3}{2\pi k^2 t^3}}\exp\left\{-\frac{3(x-x_r)^2}{2k^2 t^3}\right\},
\label{p0_RA}
\eeq
which tends to zero as $t\to\infty$ and the reset-free process is not stationary. The presence of resetting renders the process
stationary. Indeed, substituting into Eq.~\eqref{p_stat} we write
\begin{equation}
p_{st}(x|x_r)=\sqrt{\frac{3r}{2\pi k^2}} \int_0^\infty \frac{dt}{t^{3/2}}\exp\left\{-rt-\frac{3(x-x_r)^2}{2k^2 t^3}\right\}.
\label{ps_ra1}
\end{equation}

Since the integrand is continuous, positive and tends to zero exponentially (both as $t\to 0$ and $t\to \infty$) when $x\neq x_r$, the integral in Eq.~\eqref{ps_ra1} is finite and not vanishing. The complete process is thus stationary. However, the integral in Eq.~\eqref{ps_ra1} is difficult to perform analytically in exact form and results in a very lengthy expression involving Airy and Kelvin special functions which we will not reproduce here. We obtain instead an approximate expression for it using the Laplace method one more time. Let us define
\begin{equation}
\x \equiv\sqrt{\frac{3(x-x_r)^2 r^3}{2 k^2}},\qquad
\s \equiv r t \,,
\label{xi}
\end{equation}
and write Eq.~\eqref{ps_ra1} as
\begin{equation}
p_{st}(x|x_r)=\sqrt{\frac{3 r^3}{2\pi k^2}} \int_0^\infty e^{-h(\s)}\frac{d\s}{\s^{3/2}},
\label{ps_ra2}
\end{equation}
where  
\begin{equation}
h(\s)=\s+\x^2/\s^3 \,.
\label{h}
\end{equation}

\begin{figure}[htbp]
\includegraphics[width=0.6\linewidth,keepaspectratio=true]{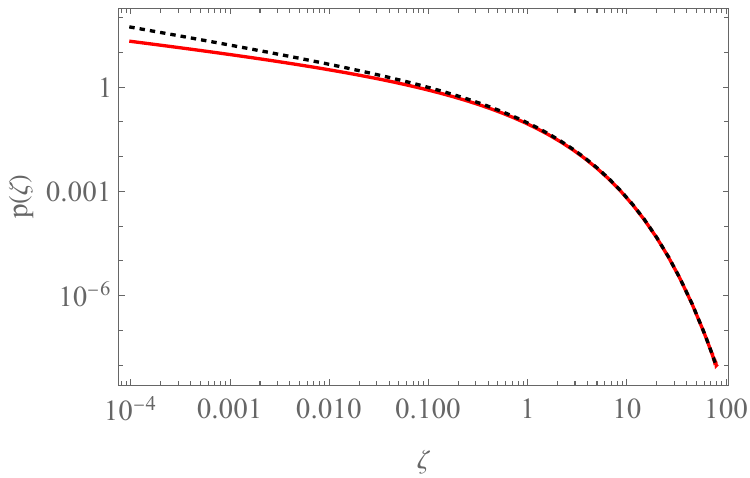}
\caption{(Color online) 
Stationary distribution for the model with random acceleration and resets. Here we show, in a double logarithmic scale, the exact stationary distribution in Eq.~\eqref{ps_ra2}, solid red line, and the approximate result in Eq.~\eqref{ps_ra3}, dotted black line. In both cases we have depicted the dimensionless quantity $p(\x)\equiv p_{st}(x|x_r) |dx/d\x|$, where $\x$ is defined in Eq.~\eqref{xi}. The concordance for $\x\gtrsim 1$ is remarkable.}  
\label{Fig:p_st_ra}
\end{figure}

The Laplace approximation (cf. Appendix \ref{app_ps}) gives
\begin{equation}
p_{st}(x|x_r)\simeq \frac{(2r^3/k^2)^{1/4}}{2|x-x_r|^{1/2}}\exp\left\{-\frac{4(r^3/2)^{1/4}}{(3k)^{1/2}}|x-x_r|^{1/2}\right\},
\label{ps_ra3}
\end{equation}
for $\x\neq 0$. Note that the integral in Eq.~\eqref{ps_ra2} does not converge for $\x=0$, due to the behavior of $\s^{-3/2}$ for $\s\to 0$, and that the goodness of the approximation will improve for increasing values of $\x$, since this will move the minimum of $h(\s)$ from away the divergence at $\s=0$, as can be observed in Fig.~\ref{Fig:p_st_ra}.

Figure~\ref{Fig:p_st_ra} also shows how the approximation given by Eq.~\eqref{ps_ra3} overestimates Eq.~\eqref{ps_ra2} for small values of $\x$. For a given model, where $r$ and $k$ are fixed, changes in $\x$ imply changes in $|x-x_r|$, resulting in the approximate stationary distribution not being normalized. This can be directly checked since the integration of Eq.~\eqref{ps_ra3} yields
$$
\int_{-\infty}^\infty p_{st}(x|x_r)dx = \sqrt{3/2} > 1.
$$
Then, if necessary, we can renormalize it by dividing the right hand side of Eq.~\eqref{ps_ra3} by $\sqrt{3/2}$  giving thus 
\begin{equation}
p_{st}(x|x_r)\simeq \frac{1}{\sqrt{6k}}\frac{(2r^3)^{1/4}}{|x-x_r|^{1/2}}\exp\left\{-\frac{4(r^3/2)^{1/4}}{(3k)^{1/2}}|x-x_r|^{1/2}\right\}.
\label{ps_ra}
\end{equation}

\subsection{Inertial Brownian motion}

\begin{figure}[htbp]
\includegraphics[width=0.6\linewidth,keepaspectratio=true]{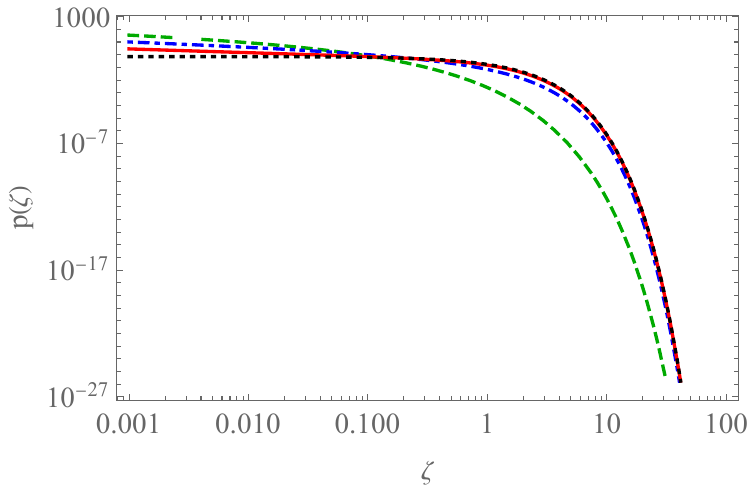}
\caption{(Color online) 
Stationary distribution for the inertial Brownian motion with resets. Here we show, in a double logarithmic scale, the numerical evaluation of the exact stationary distribution for: $\beta = r/10$, dashed green line; $\beta = r$, dot-dashed blue line; and $\beta=10\,r$, solid red line; as well as the approximate result in Eq.~\eqref{p_st_bm}, dotted black line. In all cases we have depicted the dimensionless quantity $p(\x)\equiv p_{st}(x|x_r) |dx/d\x|$, where $\x$ is defined in Eq.~\eqref{xi_bm}. 
} 
\label{Fig:p_st_bm}
\end{figure}

For the inertial Brownian motion, we can easily see from  Eq.~\eqref{asym_sigma} that that $\sigma_x(t) \to \infty$ as $t\to\infty$, implying that $p_0(x,t|x_0,y_0)\to 0$ and there is no stationary distribution for the reset-free process $p_{st}^{(0)}(x)$ (a well known result of the free Brownian motion). 
For $\beta \gg r$ 
we can use for the variance $\sigma_x^2(t)$ the limit expression given in Eq.~\eqref{asym_sigma}, and substituting the latter into 
into Eq.~\eqref{gauss_p_st} we write the approximation 
$$
p_{st}(x|x_r)\simeq \frac{\beta}{k} \sqrt{\frac{r}{2\pi}} \int_0^\infty \frac{d\s}{\s^{1/2}} \exp\left\{-\s-\frac{\x^2}{2 \s}\right\},
$$
where $\s=r t$ and 
\beq
\x\equiv \sqrt{\frac{r \beta^2(x-x_r)^2}{k^2}}.
\label{xi_bm}
\eeq
Taking into account that \cite{roberts}
$$
\int_0^\infty \frac{d\s}{\s^{1/2}}e^{-\s-a/\s}=\sqrt{\pi}e^{-2a^{1/2}}, \qquad (a>0),
$$
we obtain
\begin{equation}
p_{st}(x|x_r)\simeq \frac{\beta}{k}\left(\frac r2\right)^{1/2}\exp\left\{-\frac{2\beta}{k}\left(\frac r2\right)^{1/2}|x-x_r|\right\},
\label{p_st_bm}
\end{equation}
an expression valid for $\beta\gtrsim r$ which clearly shows that Brownian motion under resettings is stationary while the reset-free process is not. Let us finally note that this approximate form of the stationary PDF is properly normalized, since in this case we have estimated not the value of the integral in Eq.~\eqref{gauss_p_st}, but the value of the integrand. As it can be seen in Fig.~\ref{Fig:p_st_bm}, this results in a succession of curves for the PDF that steadily converge to Eq.~\eqref{p_st_bm} as $\beta \gg r$. On the other side, recall that the case $\beta \ll r$ will lead us back to the previous scenario, Eqs.~\eqref{ps_ra2} and~\eqref{ps_ra3}.

\section{Mean first-arrival time}
\label{sec_mfat}

As mentioned in Sect.~\ref{intro} a key characteristic of resetting is a potential decrease in the mean first-arrival time (MFAT). Dealing with first-passage and extreme times for inertial random processes is rather difficult \cite{maso_llibre} and only in few cases it is possible to get analytical expressions for the mean exit time \cite{mas_por_95,mas_por_96}. We here present a first and incomplete approximation on how resettings affect first-passage times of inertial processes. Suppose we are interested on the MFAT to some critical level (or threshold) $x_c$ of an inertial random process $X(t)$ under resettings. The problem is characterized by the survival probability (SP) to $x_c$ which is the probability that the process, being initially in the state $\mathbf u_0=(X(t_0),\dot X(t_0))$,  does not reach threshold $x_c$ during the time interval $[t_0,t]$. We denote this probability by $S_r(t|\mathbf u_0,t_0; \mathbf u_r,t_r)$, where $t_r$ is the last reset time prior to $t_0$ ($t_r\leq t_0$) and $\mathbf u_r=(X(t_r),\dot X(t_r))$. Let us also denote by $S_0(t|\mathbf u_0,t_0)$ the survival probability to threshold $x_c$ of the reset-free process. Both probabilities $S_r$ and $S_0$ are related to each other by a renewal equation \cite{pal_2016}. Indeed, if we assume that at $t_0$ a resetting event has occurred, so that $t_r=t_0$, then such an equation reads \cite{maso_mont_2019}
$$
S_r(t|\mathbf u_0,t_0)=e^{-rt}S_0(t|\mathbf u_0,t_0)+r\int_{t_0}^t e^{-r(t'-t_0)}S_r(t|\mathbf u_r,t')S_0(t'|\mathbf u_0,t_0) dt'.
$$
Note that the first term on the right-hand side represents the probability that neither a reset has occurred at time $t$ nor any hitting to threshold $x_c$ between $t_0$ and $t$. The second term is the probability that the first resetting (after the one at $t_0$) to position $x_r$ has occurred at some intermediate time $t'\in[t_0,t]$ with no hitting to $x_c$  between $t_0$ and $t'$ and also from $t'$ and $t$, all of this integrated over any intermediate time $t'$.

Assuming time homogeneity we may set $t_0=0$ without loss of generality and also write $S_r(t|\mathbf u_r,t')=S_r(t-t'|\mathbf u_r)$. We thus obtain the following simpler integral equation for $S_r$:
\begin{equation}
S_r(t|\mathbf u_0)=e^{-rt}S_0(t|\mathbf u_0)+r\int_{0}^t e^{-rt'}S_0(t'|\mathbf u_0)S_r(t-t'|\mathbf u_r) dt'.
\label{renewal_S}
\end{equation}

Taking the Laplace transform of this equation we have
\begin{equation}
\hat S_r(s|\mathbf u_0)=\hat S_0(r+s|\mathbf u_r)+r\hat S_0(r+s|\mathbf u_0)\hat S_r(s|\mathbf u_r),
\label{lt_1}
\end{equation}
where
$$
\hat S_r(s|\mathbf u_0)=\int_0^\infty e^{-st} S_r(t|\mathbf u_0)dt,
$$
and similarly for $\hat S_0$. Setting $\mathbf u_r=\mathbf u_0$ in Eq.~\eqref{lt_1} we obtain 
\begin{equation}
\hat S_r(s|\mathbf u_r)=\frac{\hat S_0(r+s|\mathbf u_r)}{1-r\hat S_0(r+s|\mathbf u_r)},
\label{lt_2}
\end{equation}
which relates the SP of the resetting process with the SP of the reset-free process when $\mathbf u_r=\mathbf u_0$, that is, when resettings bring the process to the initial state. In the general case when $\mathbf u_r\neq \mathbf u_0$, substituting Eq.~\eqref{lt_2} into Eq.~\eqref{lt_1} and simplifying yields~\cite{majumdar_11prl,majumdar_11jpa}: 
\begin{equation}
\hat S_r(s|\mathbf u_0)=\frac{\hat S_0(r+s|\mathbf u_0)}{1-r\hat S_0(r+s|\mathbf u_r)},
\label{lt_3}
\end{equation}

In terms of the SP the mean first-arrival time is given by the time integral \cite{redner_book,maso_llibre} 
$$
T_r(\mathbf u_0)=\int_0^\infty S_r(t|\mathbf u_0) dt=\hat S_r(s=0|\mathbf u_0),
$$
and setting $s=0$ in Eq.~\eqref{lt_3} we get
\begin{equation}
T_r(\mathbf u_0)=\frac{\hat S_0(r|\mathbf u_0)}{1-r\hat S_0(r|\mathbf u_r)}.
\label{mfat}
\end{equation}

Before proceeding further, let us incidentally note that a similar relation between the MFAT of the resetting process with the survival probability of the reset-free process can be obtained when resetting times are not exponentially distributed (i.e., they are not Poissonian) but distributed by a general probability density function $\psi(t)$. In such a case Chechkin and Sokolov \cite{sokolov_18} have recently shown that,
\begin{equation}
T_r(\mathbf u_0)=\frac{\int_0^\infty\psi(t)\int_0^tS_0(t'|\mathbf u_0)dtdt'}{\int_0^\infty \psi(t)[1-S_0(t|\mathbf u_r)]dt},
\label{mfat_gen}
\end{equation}
from which we directly recover Eq.~\eqref{mfat} in the Poissonian case when $\psi(t)=re^{-rt}$. 

Equations~\eqref{mfat} and~\eqref{mfat_gen} show that in order to obtain the MFAT for the resetting process one needs to know the complete survival probability of the reset-free process. Unfortunately obtaining $S_0(t|\mathbf u_r)$ (or its Laplace transform) for inertial processes seems to be out of reach even in the simplest case of random acceleration. In such a case we obtained some years ago \cite{mas_por_95,mas_por_96} the exact expression for the MFAT, $T_0(\mathbf u_0)$, however obtaining the survival probability $S_0(t|\mathbf u_0)$ has proved to be rather difficult, not to say impossible (see, nonetheless, the quite recent mathematical work \cite{hyung_etal_2014}). 

For this reason we will next address an analytical approximation to the MFAT valid for small frequencies of resetting. Assuming that $\hat S_0(s|\mathbf u_0)$ is a differentiable function of the Laplace variable $s$, then for small values of the resetting rate $r$ we may write
\begin{equation}
\hat S_0(r|\mathbf u_0)=\hat S_0(0|\mathbf u_0)+r \hat S'_0(0|\mathbf u_0)+O(r^2),
\label{small_r}
\end{equation}
but 
$$
\hat S'_0(0|\mathbf u_0)=-\int_0^\infty t S_0(t|\mathbf u_0) dt.
$$
Now defining
\begin{equation}
\lambda_0(\mathbf u_0)\equiv\int_0^\infty t S_0(t|\mathbf u_0) dt=-\hat S'_0(0|\mathbf u_0) \geq 0,
\label{lambda}
\end{equation}
and recalling that $\hat S_0(0|\mathbf u_0)=T_0(\mathbf u_0)$, we write Eq.~\eqref{small_r} as
$$
\hat S_0(r|\mathbf u_0)=T_0(\mathbf u_0)-r \lambda_0(\mathbf u_0)+O(r^2),
$$
which substituting into Eq.~\eqref{mfat} yields
$$
T_r(\mathbf u_0)=\frac{T_0(\mathbf u_0)-r \lambda_0(\mathbf u_0)+O(r^2)}{1-rT_0(\mathbf u_r)+O(r^2)},
$$

Expanding up to first order we have
$$
T_r(\mathbf u_0)=T_0(\mathbf u_0)\left\{1-r\left[\frac{\lambda_0(\mathbf u_0)}{T_0(\mathbf u_0)}-T_0(\mathbf u_r)\right]+O(r^2)\right\}.
$$
and within the same degree of approximation we may exponentiate and write
\begin{equation}
T_r(\mathbf u_0)=T_0(\mathbf u_0)\exp\left\{-r\left[\frac{\lambda_0(\mathbf u_0)}{T_0(\mathbf u_0)}-T_0(\mathbf u_r)\right]+O(r^2)\right\},
\label{mfat_approx}
\end{equation}
an expression associating, for small values of the resetting rate, the MFAT for the resetting process with the MFAT of the reset-free process.

Let us finally note that if  
\begin{equation}
\lambda_0(\mathbf u_0)\geq T_0(\mathbf u_0)T_0(\mathbf u_r),
\label{Tr_dec}
\end{equation}
then 
$$
T_r(\mathbf u_0)\leq T_0(\mathbf u_0),
$$
and resettings may decrease the MFAT (at least for small resetting rates, $r\to 0$). 

In fact, for $r\to \infty$ one has $T_r(\mathbf u_0)\to \infty$, since then the process will reach $\mathbf u_r$ instantly, and be trapped there. Therefore, if Eq.~\eqref{Tr_dec} holds, $T_r(\mathbf u_0)$ must have (at least) a minimum,
\begin{equation}
\left. \frac{\partial}{\partial r} T_r(\mathbf u_0)\right|_{r=r^*}=0.
\label{Tr_min}
\end{equation}
When the  optimal value $r^*$ is small, the minimum can be estimated as follows. Expanding $\hat S_0(r|\mathbf u_0)$ we have
$$
\hat S_0(r|\mathbf u_0)=T_0(\mathbf u_0)-r \lambda_0(\mathbf u_0)+\frac{1}{2} r^2 \nu_0(\mathbf u_0) + O(r^3),
$$
where
\begin{equation}
\nu_0(\mathbf u_0)\equiv\int_0^\infty t^2 S_0(t|\mathbf u_0) dt.
\label{nu}
\end{equation}
Substituting for Eq.~\eqref{mfat} we have
\begin{equation}
T_r(\mathbf u_0)=\frac{T_0(\mathbf u_0)-r \lambda_0(\mathbf u_0)+\frac{1}{2} r^2 \nu_0(\mathbf u_0)+ O(r^3)}{1-rT_0(\mathbf u_r)+r^2 \lambda_0(\mathbf u_r)+O(r^3)},
\label{mfat_r}
\end{equation}
and from Eq.~\eqref{Tr_min} we obtain the approximate expression for the optimal resetting rate,
\begin{equation}
r^*\simeq \frac{\lambda_0(\mathbf u_0)- T_0(\mathbf u_0)T_0(\mathbf u_r)}{\nu_0(\mathbf u_0) - 2T_0(\mathbf u_r) \lambda_0(\mathbf u_0) - 2 T_0(\mathbf u_0)\left[\lambda_0(\mathbf u_r)-T_0^2(\mathbf u_r) \right]}.
\label{r*}
\end{equation}
The numerator of this expression is positive because of condition in Eq.~\eqref{Tr_dec}. Thus $r^*$ will be meaningful (i.e., positive) provided that the denominator is also positive, that is 
\begin{eqnarray}
\nu_0(\mathbf u_0) &>& 2T_0(\mathbf u_r) \lambda_0(\mathbf u_0) + 2 T_0(\mathbf u_0)\left[\lambda_0(\mathbf u_r)-T_0^2(\mathbf u_r) \right]
\label{r*_condition}
\\
&=&2 T_0(\mathbf u_0) \lambda_0(\mathbf u_r) + 2 T_0(\mathbf u_r)\left[\lambda_0(\mathbf u_0)-T_0(\mathbf u_0)T_0(\mathbf u_r) \right],
\nonumber
\end{eqnarray}
being both summands positive-definite (cf. Eq.~\eqref{Tr_dec}) and, hence, $\nu_0(\mathbf u_0)>0$. 
 
Let us incidentally note that for a deterministic process, $X(t)=x_0+y_0 t$, with $x_c>x_0$ and $y_r=y_0=v>0$, in this case Eq.~\eqref{Tr_dec} implies that 
$x_r\geq (x_c+x_0)/2$ and
$$
r^*\simeq \frac{3v}{2} \frac{(x_c-x_0)\left[2 x_r-x_c-x_0\right] }{(x_c-x_r)^3+ (x_r-x_0)^3} >0.
$$

We therefore come to the conclusion that if the (reset-free) survival first moment, Eq.~\eqref{lambda}, satisfies the condition in Eq.~\eqref{Tr_dec} resettings may decrease the MFAT. In such a case there may exists an optimal rate $r^*$ for which the MFAT is minimum. For small resetting rates, such an optimal rate is approximately given by Eq.~\eqref{r*} as far as the (reset-free) survival second moment, Eq.~\eqref{nu}, satisfies the condition in Eq.~\eqref{r*_condition}.

\section{Concluding remarks}
\label{conclusions}

We have studied the effect of resetting events on the counting of level crossings for linear inertial processes driven by Gaussian white noise. 
After the brief review in Sect. \ref{sec_general} of the work recently published by two of us \cite{maso_pala} on Rice's theory of level counting and its generalization to inertial processes, we have shown that for inertial processes resettings affect both position and velocity and have obtained the equations relating the joint PDF (position and velocity) of the reset-free process with that of the complete resetting process.  

We have next addressed the main objective of the paper which is to determine the effect of stochastic resettings on the counting of crossing events. We have thus obtained, for Poissonian resettings of rate $r$, renewal equations that allow us to obtain the crossing intensity $\mu_u(t)$ knowing the intensity $\mu_u^{(0)}(t)$ of the underlying reset-free process, cf. Eq.~\eqref{ie_mu1},
$$
\mu_u  (t|x_0,y_0)=e^{-rt}\mu_u^{(0)}(t|x_0,y_0)+r\int_{0}^t e^{-rt'} \mu_u^{(0)}(t'|x_r,y_r) dt'.
$$
From this fundamental expression we see that a nonzero stationary crossing intensity $\mu_u = \lim_{t\to\infty} \mu_u(t)$ is simply given by the Laplace transform of the reset-free intensity, 
$$
\mu_u=r\hat\mu_u^{(0)}(r|x_r,y_r),
$$
which shows that that stationary crossing intensity can be nonzero 
even if the intensity of the reset-free process vanishes as $t\to\infty$.

We have applied these results to two simple, but relevant, examples: the random acceleration process and the inertial Brownian motion. Both processes are not stationary and their reset-free crossing intensity vanishes at large times. However, as we have just mentioned, in the presence of resettings a nonzero $\mu_u$ exists. Even more relevant is that in both cases, we found that  $\mu_u$ displays a maximum as a function of $r$, due to the fact that as $r\to 0$ the process wanders away from any finite level $u$, while as $r\to \infty$ it does not have the time to undergo a crossing between resetting events. We have thus shown that, at least for these two cases, resettings increase the stationary frequency of crossing events. Very likely this remarkable fact is also true for any inertial process under stochastic resetting, which calls for further investigation and  is relevant for applications 
in which one wishes to minimize the return time of a process to a given level. 

For the random acceleration process, $\mu_u$ depends on three timescales: $r^{-1}$, $\tau_u \sim u^{2/3}$, and $\tau_1 \sim y_0^2$, where the last two are intrinsic timescales of the reset-free process representing, respectively, the time to reach a level $u$ starting from a zero level and zero velocity, and the time to reach a velocity $y_0$ starting from zero velocity.

We have studied in detail the scaling behavior of $\mu_u$ in two special cases in which either $u$ or $y_0$ are set to zero (as $\mu_u$ diverges when they are both zero), 
and obtained asymptotic expressions for small and large values of $r \tau_u$ and $r \tau_1$. For $y_0 = 0$, $u\neq 0$, to leading order $\mu_u$ goes to zero as a stretched exponential for large $r \tau_u$,
\bd 
\mu_u  \, \, \underset{r \tau_u \to \infty}{\simeq} \,\, a \, r \, \Exp \left[-b (r \tau_u)^{3/4}\right],
\ed
where $a$ and $b$ are numerical constants |see Eq.~\eqref{large_r_RA_y0}. In the opposite limit as $\tau\tau_u \to 0$, $\mu_u$ diverges logarithmically to leading order (cf. Eq.~\eqref{small_r_RA_y0})  as
\bd
\mu_u \,\,\underset{r \tau_u \to 0}{\simeq}\,\, - \frac{\sqrt{3}}{2 \pi} r \ln (r\tau_u) \,.
\ed 
For $u = 0$, $y_0\neq 0$ we obtain similar behaviors to leading order:
\bd
\mu_0  \,\, \underset{r \tau_1 \to \infty}{\simeq}  \,\, \frac{1}{2} r e^{-\sqrt{6 r \tau_1}}, \qquad \mu_0 \,\, \underset{r \tau_1 \to 0}{\simeq}\,\, -\frac{\sqrt{3}}{2 \pi} r \ln (r \tau_1).
\ed 

In both cases, we also carried out an analogous study of the stationary upcrossing and downcrossing intensities. One of the two always dominates the total intensity, and for the subdominant one we find similar asymptotic expressions to those given above for the total intensity.

In the inertial Brownian motion a fourth timescale appears, namely, the inverse of the damping constant, $\beta^{-1}$. In this case we have studied only the case of zero initial velocity. The above large $r$ expression for random acceleration becomes asymptotically exact for inertial Brownian motion in the regimes  $r^{-1} \tau_u \to \infty$ and 
$\beta \tau_u \to 0$. For small $r$, we find the simple and rather appealing expression 
\bd
\mu_u \,\,\underset{r /\beta \to 0}{\simeq}\,\, 
\left(\frac{\beta r}{2\pi}\right)^{1/2},
\ed
independently of the level $u$.

Analogously to non-inertial, first-order, processes, resettings stabilize inertial processes in the sense that nonstationary processes become stationary as soon as the resetting mechanism has been set in.  We have thus been able to obtain approximate  explicit expressions for the stationary distribution of both the random acceleration process (cf. Eq.~\eqref{ps_ra}) and the inertial Brownian motion (cf. Eq.~\eqref{p_st_bm}).  

We have finally discussed the effect of resettings on the MFAT of inertial processes.  We have reviewed the question and showed that the MFAT of the complete resetting process $T_r(x_0,y_0)$ is related to the survival probability $S_0(t|x_0,y_0)$ of the reset-free process which, for Poissonian resettings, is given by the simple relation (cf. Eq.~\eqref{mfat})
$$
T_r(x_0,y_0)=\frac{\hat S_0(r|x_0,y_0)}{1-r\hat S_0(r|x_r,y_r)},
$$
where $\hat S_0(r|x_0,y_0)$ is the Laplace transform of the rest-free survival probability. From a practical point of view, the problem with this simple expression lies on the difficulty of obtaining (even approximate forms) of $S_0(t|x_0,y_0)$. We could, however, obtain an approximate expression, valid for small resetting rates $r$,  relating $T_r(x_0,y_0)$ with the MFAT of the underlying reset-free process $T_0(x_0,y_0)$  (cf. Eq.~\eqref{mfat_approx}) 
$$
T_r(x_0,y_0)=T_0(x_0,y_0)\exp\left\{-r\left[\frac{\lambda_0(x_0,y_0)}{T_0(x_0,y_0)}-T_0(x_r,y_r)\right]+O(r^2)\right\},
$$
where $\lambda_0$ is defined in Eq.~\eqref{lambda}. Finally, if  $\lambda_0(x_0,y_0)\geq T_0(x_0,y_0)T_0(x_r,y_r)$, then $T_r(x_0,y_0)\leq T_0(x_r,y_r),$ 
and, as in non inertial processes, resettings may reduce the MFAT.

Let us finally remark that the present approach can be extended to include processes driven by Gaussian colored noise. This work is under current research and some results will be published soon.

\acknowledgments 

This work has been partially funded by MCIN/AEI/10.13039/501100011033 and by ``ERDF A way of making Europe", grant numbers PID2022-140757NB-I00 (M.M. and J.M.), PID2022-39913NB-I00 (M.P.), and by Generalitat de Catalunya, grant numbers 2021SGR00856 (M.M. and J.M.), 2021SGR00247 (M.P.). M.P. thanks Marco Palassini Vidal for inspiration.

\appendix
\section{Approximate expressions for Eq.~\eqref{scaling_RA_y0}}
\label{app_f}

For large $s$, the most important contribution to $f(s)$,
\beq
f(s) \equiv \frac{s \sqrt{3}}{2 \pi} \int_0^\infty \frac{1}{\theta}\left[e^{-\frac{9}{2 \theta^3}}+ 3\sqrt{\frac{\pi}{2\theta^3}} 
\mbox{Erf}\left(\frac{3}{\sqrt{2 \theta^3}}\right)\right] e^{-s \theta -\frac{3}{2 \theta^3}} d\theta ,
\label{scaling_RA_y0_app}
\eeq 
comes from those values of $\theta$ in the vicinity of $\theta_m$, the single 
minimum of the function
$$
h(\theta)=s \theta +\frac{3}{2 \theta^3},
$$
that is 
$$
\theta_m=\left(\frac{9}{2 s}\right)^{\frac{1}{4}}\, .
$$
We can then  use the Laplace method \cite{erderlyi} to evaluate the
above integral, and since $\theta_m$ becomes very small when $s$ is large,
we can  neglect the first term inside the square brackets in Eq.~\eqref{scaling_RA_y0_app}. This gives
\begin{eqnarray*}
f(s) \underset{s\to \infty} \simeq f_{\gg}(s) &\equiv& \frac{s \sqrt{3}}{2 \pi} \sqrt{\frac{2 \pi}{h^{''}(\theta_m)}}  \frac{1}{\theta_m} 3 \sqrt{\frac{\pi}{2\theta_m^3}} \mbox{Erf}\left(\frac{3}{\sqrt{2 \theta_m^3}}\right) e^{-h(\theta_m)} \\
&=&  \frac{s}{2} \sqrt{\frac{3}{2}} \Erf\left(\frac{3^{1/4}}{2^{1/8}} s^{3/8}\right) \mbox{Exp}\left(-\frac{2^{7/4}}{\sqrt{3}} s^{3/4}\right),
\end{eqnarray*}
where we used
$$
h(\theta_m)=\frac{4}{3} s \theta_m= \frac{2^{7/4}}{\sqrt{3}} s^{3/4}, \quad
h''(\theta_m)=\frac{18}{\theta_m^5}.
$$
When $s$ is small enough, we can no longer ignore the contribution of the first term inside the square brackets in Eq.~\eqref{scaling_RA_y0_app}. Let us define 
$$
f_1(s) \equiv\frac{s \sqrt{3}}{2 \pi} \int_0^\infty \frac{1}{\theta} e^{-s \theta -\frac{6}{\theta^3}} d\theta,
$$
and
\beq
f_2(s) \equiv \frac{s 3\sqrt{3}}{2 \sqrt{2 \pi}} \int_0^\infty  \frac{1}{\sqrt{\theta^5}}
\mbox{Erf}\left(\frac{3}{\sqrt{2 \theta^3}}\right) e^{-s \theta -\frac{3}{2 \theta^3}} d\theta, 
\label{f2_s_small}
\eeq
such that $f(s)=f_1(s)+f_2(s)$. By analyzing $f_1(s)$ one concludes that the integrand is not sharply peaked around a given value. On the contrary, it abruply attains a maximum and then decays exponentially slowly. Let us denote by $\theta_*$ the location of the maximum of the whole integrand and split the integral in Eq.~\eqref{scaling_RA_y0_app} into two terms,  $f_{1,1}(s)$ and $f_{1,2}(s) $,
\begin{eqnarray*}
f_{1,1}(s) &\equiv& \frac{s \sqrt{3}}{2 \pi} \int_0^{\theta_*} \frac{1}{\theta} e^{-s \theta -\frac{6}{\theta^3}} d\theta=  \frac{s \sqrt{3}}{2 \pi} \int_0^{\theta_*} e^{-s \theta -\frac{6}{\theta^3}-\ln(\theta)} d\theta,\\
f_{1,2}(s) &\equiv& \frac{s \sqrt{3}}{2 \pi} \int_{\theta_*}^{\infty} \ \frac{1}{\theta} e^{-s \theta -\frac{6}{\theta^3}}  d\theta .
\end{eqnarray*}
When $s$ is small enough, the value of $\theta_*$ can be estimated through the minimum of 
$$
h(\theta)=\frac{6}{\theta^3}+\ln(\theta),
$$
that is $\theta_*\simeq18^{1/3}$, and therefore 
$$
f_{1,1}(s) \simeq \frac{1}{2}  \frac{s \sqrt{3}}{2 \pi} \sqrt{\frac{2 \pi}{h^{''}(\theta_*)}}  e^{-s \theta_*} e^{ -h(\theta_*)}=\frac{s }{2\sqrt{2 \pi}}  e^{-s 18^{1/3}-1/3}  \simeq \frac{s }{2\sqrt{2 \pi}} e^{-1/3} ,
$$
where we have used again the Laplace method and the $1/2$ pre-factor is due to the fact that we have just one side of the peak. The integral for values $\theta >\theta_*$ has the upper bound
$$
f_{1,2}(s) < \frac{s \sqrt{3}}{2 \pi} \int_{\theta_*}^{\infty} \ \frac{1}{\theta} e^{-s \theta} d\theta = \frac{s \sqrt{3}}{2 \pi}  \Gamma(0,s\theta_*)\simeq -\frac{s \sqrt{3}}{2 \pi} \left[\gamma + \ln\left(s 18^{1/3}\right)\right],
$$
where $\Gamma(a,u)$ is the upper incomplete Gamma function. In order to compute $f_2(s)$ when $s\to 0$ one can perform a change of the variable of integration in Eq.~\eqref{f2_s_small} in such a way the argument of the error function becomes the new variable and the exponential is expanded up to order zero:
$$
f_2(s) = \frac{s}{\sqrt{3 \pi}} \int_0^\infty  \mbox{Erf}\left(z\right) e^{-\frac{z^2}{3}} \Exp\left( -s \sqrt[3]{\frac{9}{2 z^2}}\right)dz \simeq \frac{s}{\sqrt{3 \pi}} \int_0^\infty  \mbox{Erf}\left(z\right) e^{-\frac{z^2}{3}} dz=\frac{s}{3 }.
$$

Collecting all the terms we finally have
$$
f_{\ll}(s) \equiv \frac{s \sqrt{3}}{2 \pi} \left[\sqrt{\frac{\pi}{6}} e^{-1/3}-\gamma - \ln(s 18^{1/3})\right] + \frac{s}{3}.
$$

\section{Approximate expressions for Eq.~\eqref{mupm_RA_y0}}
\label{app_fm}

We want to derive approximate expressions for
 \beq
f^{(-)}(s)=\frac{s \sqrt{3}}{4 \pi}
 \int_0^\infty \frac{1}{\theta}\left[e^{-\frac{9}{2 \theta^3}} - 3 \,\sgn(u) \sqrt{\frac{\pi}{2\theta^3}} 
\mbox{Erfc}\left(\frac{3}{\sqrt{2 \theta^3}}\right)\right] e^{-s \theta  -\frac{3}{2 \theta^3}} d\theta ,
\label{mupm_RA_y0_app}
\eeq
for $u>0$ when $s$ is either large or small. In this case, in contrast to the approach taken in Appendix~\ref{app_f}, one must consider the whole expression inside the square brackets at once, otherwise one may get inconsistencies. After a change of variables, $f^{(-)}(s)$ reads
 \beq
f^{(-)}(s)=
\frac{s }{2 \pi \sqrt{3}}
 \int_0^\infty \frac{1}{z} \left[e^{-z^2} - \sqrt{\pi}\, z \,
\mbox{Erfc}\left(z\right)\right] e^{-s  \sqrt[3]{\frac{9}{2 z^2}}  -\frac{z^2}{3}} dz .
\label{mupm_RA_y0_app2}
\eeq
When $s\to \infty$, the most important contribution to the integral will come from large values of $z$. Using the asymptotic series
$$
\mbox{Erfc}\left(z\right)=\frac{e^{-z^2}}{ \sqrt{\pi}}  \sum_{n=0}^{\infty}(-1)^n\frac{(2 n)!}{n! (2 z)^{2 n+1}},
$$
we obtain
$$
e^{-z^2} - \sqrt{\pi}\, z \,\mbox{Erfc}\left(z\right)\simeq \frac{e^{-z^2}}{2 z^2}, 
$$
hence
\beq
f^{(-)}(s)\simeq
\frac{s }{4 \pi \sqrt{3}}
 \int_0^\infty \frac{1}{z^3}  e^{-s  \sqrt[3]{\frac{9}{2 z^2}}  -\frac{4 z^2}{3}} dz .
\label{mupm_RA_y0_app3}
\eeq
This integral can be approximated once again by the Laplace method to obtain
$$
f^{(-)}(s) \underset{s\to \infty} \simeq f^{(-)}_{\gg}(s)=\frac{1}{\sqrt{\pi} (2^{15} 9 s)^{1/8}}\Exp\left(-\frac{2^{9/4}}{\sqrt{3}} s^{3/4} \right).
$$
For $s\ll 1$ we must follow an approach similar to that in Appendix \ref{app_f}, with the main difference concentrated in the fact that in this case all the expression within the square brackets in the integrand of Eq.~\eqref{mupm_RA_y0_app} is approximately equal to one since when $z\to 0$:
$$
e^{-z^2} - \sqrt{\pi}\, z \,\mbox{Erfc}\left(z\right)\simeq 1.
$$
Thus, for $s\ll 1$, 
\beq
f^{(-)}(s) \simeq
\frac{s \sqrt{3}}{4 \pi}
 \int_0^\infty \frac{1}{\theta} e^{-s \theta  -\frac{3}{2 \theta^3}} d\theta .
\label{mupm_RA_y0_app4}
\eeq
Again, we must split the integration range in Eq.~\eqref{mupm_RA_y0_app4} in two zones,
\begin{eqnarray*}
f^{(-)}_{1}(s) &\equiv& \frac{s \sqrt{3}}{4 \pi} \int_0^{\theta_*} \frac{1}{\theta} e^{-s \theta -\frac{3}{2 \theta^3}} d\theta=  \frac{s \sqrt{3}}{4 \pi} \int_0^{\theta_*} e^{-s \theta -\frac{3}{2 \theta^3}-\ln(\theta)} d\theta,\\
f^{(-)}_{2}(s) &\equiv& \frac{s \sqrt{3}}{4 \pi} \int_{\theta_*}^{\infty} \frac{1}{\theta} e^{-s \theta -\frac{3}{2 \theta^3}} d\theta .
\end{eqnarray*}
When $s$ is small enough, the value of $\theta_*$ can be estimated through the minimum of 
$$
h(\theta)=\frac{3}{2\theta^3}+\ln(\theta),
$$
that is $\theta_*\simeq(9/2)^{1/3}$, and therefore 
$$
f^{(-)}_{1}(s) \simeq \frac{1}{2}  \frac{s \sqrt{3}}{4 \pi} \sqrt{\frac{2 \pi}{h^{''}(\theta_*)}}  e^{-s \theta_*} e^{ -h(\theta_*)}=\frac{s }{4\sqrt{2 \pi}}  e^{-s (9/2)^{1/3}-1/3}  \simeq \frac{s }{4\sqrt{2 \pi}} e^{-1/3} ,
$$
where we have used again the Laplace method and the $1/2$ pre-factor is due to the fact that we have just one side of the peak. The integral for values $\theta >\theta_*$ has the upper bound
$$
f^{(-)}_{2}(s) < \frac{s \sqrt{3}}{4 \pi} \int_{\theta_*}^{\infty} \ \frac{1}{\theta} e^{-s \theta} d\theta = \frac{s \sqrt{3}}{4 \pi}  \Gamma(0,s\theta_*)\simeq -\frac{s \sqrt{3}}{4 \pi} \left[\gamma + \ln\left(s  \sqrt[3]{9/2}\right)\right],
$$
where $\Gamma(a,u)$ is as above the upper incomplete Gamma function. Collecting the two terms one has finally
$$
f^{(-)}_{\ll}(s) \equiv \frac{s \sqrt{3}}{4 \pi} \left[\sqrt{\frac{\pi}{6}} e^{-1/3}-\gamma - \ln(s  \sqrt[3]{9/2})\right].
$$

\section{Derivation of Eq.~\eqref{mu_ra_exact_main}}
\label{app_g_exact}

The starting point is Eq.~\eqref{mu_ra1}:
\begin{equation}
g(s)=\frac{s \sqrt{3}}{2\pi}\int_0^\infty \frac{1}{\theta}\left[e^{-1/\theta}+ \sqrt{\frac{\pi}{\theta}} {\rm Erf} \left(1/\sqrt{\theta}\right) \right]e^{-3/\theta-s  \theta/2} d\theta.
\label{mu_ra1_app}
\end{equation}
Note that the first of the two terms in square brackets leads to \cite{roberts}
$$
\int_0^\infty \frac{1}{\theta}  e^{-4/\theta-s  \theta/2} d\theta =2 K_0\left(2\sqrt{2 s }\right),
$$
where $K_0(z)$ is a modified Bessel function of second kind \cite{mos}. As to the second term in Eq.~\eqref{mu_ra1_app} we need to expand the error function in power series,
$$
 {\rm Erf}(z) = \frac{2}{\sqrt{\pi}} \sum_{n=0}^{\infty} \frac{(-1)^n}{n!}\frac{z^{2 n+1}}{2n +1},
$$
 and therefore
 \begin{eqnarray*}
 \int_0^\infty \frac{1}{\theta}\sqrt{\frac{\pi}{\theta}} {\rm Erf} \left(1/\sqrt{\theta}\right)e^{-3/\theta-s  \theta/2} d\theta&=& 2  \sum_{n=0}^{\infty} \frac{(-1)^n}{n!}\frac{1}{2n +1} \int_0^\infty \frac{1}{\theta^{n+2}}  e^{-3/\theta-s  \theta/2} d\theta\\
 &=& 4 \sum_{n=0}^{\infty} \frac{(-1)^n}{n!}\frac{1}{2n +1} \left(\frac{s }{6}\right)^{\frac{n+1}{2}}  K_{n+1}\left(\sqrt{6 s }\right).
\end{eqnarray*}
Thus, after collecting all the terms one obtains
\begin{equation}
g(s)=\frac{s \sqrt{3}}{\pi} \left[ K_0\left(2\sqrt{2 s }\right)+ 2 \sum_{n=0}^{\infty} \frac{(-1)^n}{n!}\frac{1}{2n +1} \left(\frac{s }{6}\right)^{\frac{n+1}{2}}  K_{n+1}\left(\sqrt{6 s }\right) \right].
\label{mu_ra_exact}
\end{equation}

For small values of their arguments, the modified Bessel functions of the second kind behave as \cite{mos} 
$$
K_0(z) \simeq -\ln\left(\frac{z}{2}\right) -\gamma, \qquad\qquad K_{n}(z) \simeq \frac{(n-1)!}{2}  \left(\frac{2}{z}\right)^{n}, \quad (n\neq 0).
$$
where $\gamma$ is the Euler-Mascheroni constant, $\gamma \approx 0.5772$. Then, when $s \ll 1$ one has
\begin{equation}
g(s) \underset{s\to 0} \simeq g_{\ll}(s)  \equiv  \frac{\sqrt{3} s}{\pi} \left[ -\frac{1}{2}\ln(2 s )-\gamma + \sum_{n=0}^{\infty} \frac{(-1)^n}{2n +1} \left(\frac{1}{3}\right)^{n+1} \right]=\frac{\sqrt{3} s}{\pi} \left[ -\frac{1}{2}\ln(2 s )-\gamma +  \frac{\pi}{6 \sqrt{3}} \right].
\label{mu_ra_small}
\end{equation}

In the opposite situation, when $s \gg 1$, we use the asymptotic approximation \cite{mos},
\beq
K_{n}(z) \simeq \sqrt{\frac{\pi}{2 z}}  e^{-z}, \qquad (z\to\infty), 
\label{K_asym_app}
\eeq
and get
\begin{eqnarray}
g(s)&\simeq& \frac{s \sqrt{3}}{\pi} \left[\sqrt{\frac{\pi}{4\sqrt{2 s }}}  e^{-2\sqrt{2 s }} +2 \sqrt{\frac{\pi}{2\sqrt{6 s }}}  e^{-\sqrt{6 s }} \sum_{n=0}^{\infty}  \frac{(-1)^n}{n!}\frac{1}{2n +1} \left(\frac{s}{6}\right)^{\frac{n+1}{2}}  \right]\nonumber\\
&\simeq&\frac{s}{2} \text{Erf}\left(\sqrt[4]{\frac{s}{6}}\right)  e^{-\sqrt{6 s }},  \qquad (s\gg 1),
\label{mu_ra_large}
\end{eqnarray}
but $\text{Erf}(\sqrt[4]{s/6})\approx 1$ for $s \gg 1$ and thus we finally write
\begin{equation*}
g(s) \underset{s\to \infty} \simeq g_{\gg}(s) \equiv \frac{s}{2}   e^{-\sqrt{6 s }}.
\label{mu_ra_large_bis}
\end{equation*}

\section{Approximate expressions for Eq.~\eqref{mupm_RA_u0_y}}
 \label{app_ra_u0}

Using $\mbox{Erfc}(z)=1-\mbox{Erf}(z)$ and $\mbox{Erfc}(-z)=1+\mbox{Erf}(z)$, from Eqs.~\eqref{mu_ra1} and~\eqref{mupm_RA_u0_y} one can obtain
$$
g^{(\pm)}(s) = \frac{1}{2}g(s) \mp \frac{s}{4} \sqrt{\frac{3}{\pi}} \int_0^\infty \frac{d\theta}{\theta^{3/2}} e^{3/\theta - s  \theta/2},
$$
for $y_0>0$. The integral in the above equation can be performed exactly, giving
\beq
g^{(\pm)}(s)= \frac{1}{2}g(s) \mp \frac{s}{4} e^{-\sqrt{6 s }} \,     .
\label{gpm_gen_app}
\eeq
This formula is very useful for obtaining the asymptotic expressions of $g^{(\pm)}(s)$ for $s\ll 1$, since then (cf. Eq.~\eqref{mu_ra_small})
$$
g^{(\pm)}(s)\simeq \frac{\sqrt{3} s}{2 \pi} \left[ -\frac{1}{2}\ln(2 s )-\gamma +  \frac{\pi}{6 \sqrt{3}} \right] \mp \frac{s}{4},
$$
and therefore, in particular,
$$
g^{(+)}(s) \underset{s\to 0} \simeq g^{(+)}_{\ll}(s)  \equiv \frac{\sqrt{3} s}{2 \pi} \left[ -\frac{1}{2}\ln(2 s )-\gamma -  \frac{\pi}{3 \sqrt{3}} \right].
$$
 
Equation~\eqref{gpm_gen_app} is still useful when considering the asymptotic limit for $s\gg 1$ in the case of the zero-downcrossing intensity, since then
$$
g^{(-)}(s) \simeq g(s).
$$
In the case of the zero-upcrossing intensity, using Eq.~\eqref{gpm_gen_app} together with Eq.~\eqref{mu_ra_large} will give a negative result, so a different approach must be considered. The exact expression for $g^{(+)}(s)$ when $y_0>0$ reads
\begin{equation}
g^{(+)}(s)=\frac{s \sqrt{3}}{4\pi}\int_0^\infty \frac{1}{\theta}\left[e^{-1/\theta}- \sqrt{\frac{\pi}{\theta}} {\rm Erfc} \left(1/\sqrt{\theta}\right) \right]e^{-3/\theta-s  \theta/2} d\theta.
\label{mu_ra2_app}
\end{equation}
Here we need to use the following asymptotic expansion of the complementary error function:
$$
 {\rm Erfc}(z) = \frac{e^{-z^2}}{z \sqrt{\pi}} \sum_{n=0}^{\infty} (-1)^n \frac{(2n)!}{n! (2 z)^{2 n}},
$$
which turns Eq.~\eqref{mu_ra2_app} into
$$
g^{(+)}(s)=\frac{s \sqrt{3}}{4\pi} \sum_{n=1}^{\infty} (-1)^{n+1} \frac{(2n)!}{n! 2^{2 n}} \int_0^\infty \theta^{n-1} e^{-4/\theta-s  \theta/2} d\theta,
$$
and hence,
\begin{equation}
g^{(+)}(s)= \frac{s \sqrt{3}}{2\pi} \sum_{n=1}^{\infty} (-1)^{n+1} \frac{(2n)!}{n! (2 s)^{n/2}} K_{n}(2\sqrt{2 s}),
\label{mu_ra3_app}
\end{equation} 
where $K_n(z)$ are again the modified Bessel functions of the second kind, whose asymptotic behavior is, cf. Eq.~\eqref{K_asym_app}, 
$$
K_{n}(z) \simeq \sqrt{\frac{\pi}{2 z}}  e^{-z}, \qquad (z\to\infty).
$$
Thus we can conclude that the main contribution in Eq.~\eqref{mu_ra3_app} comes from the term with $n=1$,
$$
g^{(+)}(s) \underset{s\to \infty} \simeq g^{(+)}_{\gg}(s)= \sqrt{\frac{3}{\pi}}\left(\frac{s}{2^7}\right)^{\frac{1}{4}} e^{-2\sqrt{2 s}}.
$$

\section{Derivation of Eq.~\eqref{ps_ra3}}
\label{app_ps}

Let us first observe that for $\x\neq 0$  
$\lim_{\s\to 0} h(\s)=+\infty$,  $\lim_{\s\to \infty} h(\s)=+\infty$, and also that $h(\s)$ has a single global minimum at
\begin{equation}
\s_m=3^{1/4} \sqrt{\x},
\label{s_m_dist}
\end{equation}
with $0<\s_m<\infty$. Then, the major contribution to the integral comes from the values of $\s$ around the vicinity of $\s_m$.
Therefore we can use the Laplace approximation to obtain
$$
\int_0^\infty e^{-h(\s)}\frac{d\s}{\s^{3/2}} 
\simeq {\s_m^{-3/2}} \left[\frac{2 \pi}{h''(\s_m)}\right]^{1/2} e^{-h(\s_m)}, 
$$
and taking into account Eqs.~\eqref{xi},~\eqref{h} and~\eqref{s_m_dist} we finally obtain
$$
p_{st}(x|x_r)\simeq \frac{(2r^3/k^2)^{1/4}}{2|x-x_r|^{1/2}}\exp\left\{-\frac{4(r^3/2)^{1/4}}{(3k)^{1/2}}|x-x_r|^{1/2}\right\},
$$
for $\x\neq 0$.

\end{document}